\documentclass[12pt,letterpaper]{article}
\pdfoutput=1

\usepackage{graphicx,array}
\usepackage{color}
\usepackage{latexsym}
\usepackage{amsthm}
\usepackage{amsmath}
\usepackage{amssymb}
\usepackage{multirow}
\usepackage[numbers,sort&compress]{natbib}
\usepackage{pdflscape}	
\usepackage{bm}
\usepackage{slashed} 
\usepackage{mathrsfs}
\usepackage{tensor}
\usepackage{hyperref} 
\hypersetup{
    colorlinks=true,       
    linkcolor=red,          
    citecolor=blue,        
    filecolor=magenta,      
    urlcolor=blue           
}
\usepackage[all]{hypcap} 

\usepackage{myterms}

\newcommand{\PT}{\text{\sc pt}}

\newcommand{\Dhat}{\hat{D}}

\title{\bf Electroweak sphaleron with dimension-six operators}

\author{\large Xucheng Gan$^{a}$, Andrew J. Long$^{b}$, Lian-Tao Wang$^{a,b}$}
\date{\small \it 
$^a$Enrico Fermi Institute, University of Chicago, Chicago, Illinois 60637, USA \\
$^b$Kavli Institute for Cosmological Physics, University of Chicago, Chicago, Illinois 60637, USA 
}

\begin{document}

\maketitle

\setlength{\parskip}{0.2ex}

\begin{abstract}
New physics at the TeV scale can affect the dynamics of the electroweak phase transition in many ways.  
In this paper, we evaluate its impact on the rate of baryon-number violation via sphaleron transitions.  
We parametrize the effect of new physics with dimension-six operators, and we use the Newton-Kantorovich method to numerically solve the resulting equations of motion.  
Depending on the sign of the coefficient of the dimension-six operators, their presence can either increase or decrease the sphaleron energy at the level of a few percent, parametrically of order $m_W^2 / \Lambda^2$ where $\Lambda$ is the scale suppressing the dimension-six operator. 
The baryon number washout condition, typically written as $v_c / T_c > 1$, is directly proportional to the sphaleron energy, and we discuss how the presence of dimension-six operators can affect electroweak baryogenesis.  
\end{abstract}

\newpage

\section{Introduction}\label{sec:Introduction}

Electroweak baryogenesis provides a minimal and compelling scenario for testing the idea that the matter / antimatter asymmetry of the Universe arose at the electroweak phase transition \cite{Kuzmin:1985mm,Shaposhnikov:1986jp,Shaposhnikov:1987pf,Shaposhnikov:1987tw,Cohen:1990it,Cohen:1990py}.  
The Standard Model lacks the necessary ingredients for electroweak baryogenesis, and in general new physics is required if this scenario is to be successful.  
Various papers have studied the effect of these new particles and interactions on different aspects of electroweak baryogenesis: the strength of the electroweak phase transition, charge transport at the bubble wall, {\it CP}-violating interactions, bubble wall dynamics, and so on.  
In this work, we are interested in the effect of new physics on the washout of baryon number.  

In the Standard Model, baryon number is not a conserved charge, but rather the corresponding Abelian symmetry $\U{1}_B$ is anomalous under the electroweak gauge group $\SU{2}_L \times \U{1}_Y$ \cite{Hooft:1976up}.  
Consequently, baryon number can be violated by nonperturbative electroweak processes.  
In the Higgs phase at finite temperature, the relevant process is a gauge-Higgs field configuration called the electroweak sphaleron \cite{Manton:1983nd,Klinkhamer:1984di}.  
As particles in the plasma experience baryon-number-violating interactions, the system passes through the sphaleron configuration.  
Thus the rate of baryon-number violation is Boltzmann suppressed as $\Gamma_{\slashed{B}} \sim {\rm exp}[-E_{\rm sph}/T]$ where $E_{\rm sph}$ is the energy of the sphaleron configuration, which corresponds to the ``height'' of the potential barrier \cite{Arnold:1987mh,Arnold:1987zg,Khlebnikov:1988sr,Mottola:1990bz}.  

An accurate calculation of the sphaleron energy is important for assessing the viability of electroweak baryogenesis.  
In this scenario the matter / antimatter asymmetry of the Universe is generated during the electroweak phase transition.  
If the electroweak sphaleron is in equilibrium during the phase transition, then it will wash out the baryon asymmetry.  
The requirement of washout avoidance is expressed as a lower bound on the thermal expectation value of the Higgs field $v(T) = \langle H \rangle$.  
The washout avoidance condition is written as \cite{Kuzmin:1985mm,Shaposhnikov:1986jp,Shaposhnikov:1987pf,Shaposhnikov:1987tw}\footnote{When defined in the conventional way, the electroweak order parameter, $v(T_\PT) / T_\PT$, depends on the gauge-fixing scheme that is used to calculate the thermal effective potential \cite{Patel:2011th}.  This issue needs to be addressed in a full finite-temperature calculation of baryon number washout, but since we are primarily interested in the gauge-invariant sphaleron energy $E_{{\rm sph},0}$, we do not discuss this subtlety further here.  See also the discussion in \aref{app:Washout}.}
\begin{align}\label{intro:washout}
	\frac{v(T_{\PT})}{T_{\PT}} \gtrsim 1 \times \left( \frac{E_{{\rm sph}, 0}}{9 \TeV} \right)^{-1}
\end{align}
where $E_{{\rm sph},0}$ is the energy of the electroweak sphaleron at zero temperature, and $T_\PT \sim 100 \GeV$ is the temperature of the electroweak phase transition.  
If the presence of new physics changes the energy of the sphaleron configuration, then the washout avoidance condition is tightened or loosened accordingly \cite{Ahriche:2007jp,Funakubo:2009eg,Fuyuto:2014yia,Fuyuto:2015jha}.  

In order to remain general, we now adopt the perspective of effective field theory (EFT).  
That is, we will suppose that the presence of new particles and interactions can be captured by extending the Standard Model with operators that are built from Standard Model fields and have mass dimension $>4$.  
In fact, the leading-order operators that directly affect the sphaleron energy first appear at mass dimension-six.  
(For instance, some of these operators can arise when the Standard Model is extended by a scalar singlet field that is heavy and integrated out of the theory \cite{Buchalla:2016bse}.)  
We calculate the effect of these dimension-six operators on the sphaleron energy and thus the washout avoidance condition \pref{intro:washout}.  
Since the dimension-six operators are suppressed by $\Lambda^2$ where $\Lambda$ is the cutoff of the EFT, we can anticipate that these operators will lead to a fraction shift in the sphaleron energy on the order of $\Delta E_{{\rm sph},0} / E_{{\rm sph},0}^\SM = O(m_W^2 / \Lambda^2)$ where $m_W$ is the W-boson mass.  
Since constraints on the dimension-six operators typically require $\Lambda \gtrsim 1 \TeV$, we expect an $O(1\%)$ effect on the sphaleron energy, which is confirmed by our full numerical study.  

We have attempted to be systematic and exhaustive in our survey of dimension-six operators and their effects on the electroweak sphaleron.  
In this regard, our work can be viewed as an extension of the study in \rref{Spannowsky:2016ile} where a few higher-dimension operators were considered and the sphaleron solution was calculated.  
However, although we are motivated by baryon number washout in electroweak baryogenesis, the operators that we study here do not represent viable models of baryogenesis in and of themselves.  
This is because electroweak baryogenesis also requires that the electroweak phase transition be first order, and it requires an additional source of {\it CP} violation.  
The operator $\Ocal_6 = - (H^\dagger H)^3$ does cause the electroweak phase transition to become first order if the operator coefficient is sufficiently large \cite{Grojean:2004xa,Delaunay:2007wb}, but it does not lead to additional {\it CP} violation.  
Other operators do violate {\it CP}, but they do not affect the nature of the electroweak phase transition.  
However, electroweak baryogenesis becomes viable with a combination of multiple operators (possibly also involving the fermions, which we have not included in our study) \cite{Zhang:1993vh,Huang:2015bta,Huang:2015izx,Damgaard:2015con,Kobakhidze:2015xlz,Balazs:2016yvi}.  
Whereas it is customary to apply the washout avoidance condition as simply $v(T_\PT)/T_\PT \gtrsim 1$, our work demonstrates a degree of model dependence in the ``1.'' 

The remainder of this article is organized as follows.  
In \sref{sec:Dim6} we introduce the dimension-six operators, calculate their contribution to the Higgs and gauge field equations, and impose the sphaleron {\it Ansatz} to derive the sphaleron equations of motion.  
In \sref{sec:Sphaleron} we solve the sphaleron equations of motion numerically, and we present our main results.  
We summarize in \sref{sec:Conclusion} and propose directions for future work.  
The article is extended by four appendixes: in \aref{app:EM_Tensor} we present the energy-momentum tensor of the dimension-six operators, in \aref{app:Ansatz} we motivate the sphaleron {\it Ansatz}, in \aref{app:NK_Method} we explain the numerical technique (Newton-Kantorovich method) that was used to solve the sphaleron equations of motion, and in \aref{app:Washout} we review the derivation of the baryon number washout avoidance condition.  

\section{Dimension-Six Extension of the Electroweak Sector}\label{sec:Dim6}

\subsection{Extended Lagrangian}\label{sub:Lagrangian}

We let the Standard Model Lagrangian be extended by dimension-six operators.  
For the electroweak sphaleron solution, all of the SM fields are vanishing except for the Higgs doublet $H$, the $\SU{2}_L$ isospin gauge field $W_{\mu}^{a}$, and the $\U{1}_Y$ hypercharge gauge field $B_{\mu}$.  
Then we only need to retain the operators that are built from these fields.  
The relevant terms from the Lagrangian are written as\footnote{We use $c_i$ to denote all of the operator coefficients, whereas \rref{Elias-Miro:2013mua} uses $c_i$ for some operators and $\kappa_i$ for others.}  
\begin{align}\label{eq:L_def}
	\Lcal & = \bigl( D_{\mu} H \bigr)^{\dagger} \bigl( D^{\mu} H \bigr) - \frac{1}{4} W_{\mu \nu}^{a} W^{a \mu \nu} - \frac{1}{4} B_{\mu \nu} B^{\mu \nu} - \mu^2 H^{\dagger} H - \lambda \bigl( H^{\dagger} H \bigr)^2 + \sum_i \frac{c_i}{\Lambda^2} \, \Ocal_i 
\end{align}
where $\Lambda$ is the energy scale of new physics that suppresses the dimension-six operators $\Ocal_i$ (see below), and $c_i$ is a dimensionless coefficient.  
The Higgs covariant derivative is written as $D_{\mu} H = \bigl( \partial_{\mu} - i g \frac{\sigma^a}{2} W_{\mu}^{a} - i g^{\prime} \frac{1}{2} B_{\mu} \bigr) H$ where $g$ is the $\SU{2}_L$ isospin gauge coupling, $g^{\prime}$ is the $\U{1}_Y$ hypercharge gauge coupling, and $\sigma^a$ are the Pauli matrices.  
The isospin field strength tensor is written as $W_{\mu \nu}^{a} = \partial_\mu W_\nu^a - \partial_\nu W_\mu^a + g \epsilon^{abc} W_\mu^b W_\nu^c$ with $\epsilon^{123} = +1$, and the hypercharge field strength tensor is written as $B_{\mu \nu} = \partial_{\mu} B_{\nu} - \partial_{\nu} B_{\mu}$.  
The Higgs mass parameter is denoted by $\mu^2$, and the Higgs self-coupling is denoted by $\lambda$.  

A complete list of the dimension-six operators that are built from the SM field content appears in \rref{Elias-Miro:2013mua}\footnote{In \rref{Elias-Miro:2013mua}, the operator $\Ocal_6$ is normalized as $\Ocal_6 = \lambda \bigl( H^{\dagger} H \bigr)^3$.}.  
There are four operators involving the Higgs field,
\begin{subequations}\label{eq:op_H}
\begin{align}
	\Ocal_H & = \frac{1}{2} \bigl( \partial^{\mu} (H^{\dagger} H) \bigr)^2 \\
	\Ocal_T & = \frac{1}{2} \bigl( H^{\dagger} \overset\leftrightarrow{D^{\mu}} H \bigr)^2 \\
	\Ocal_r & = H^{\dagger} H \, \bigl( D_{\mu} H \bigr)^{\dagger} \bigl( D^{\mu} H \bigr) \\
	\Ocal_6 & = - \bigl( H^{\dagger} H \bigr)^3 
\end{align}
\end{subequations}
where $H^{\dagger} \overset\leftrightarrow{D^{\mu}} H = H^{\dagger} D^{\mu} H - \bigl( D^{\mu} H \bigr)^{\dagger} H$ and $H^{\dagger} \sigma^a \overset\leftrightarrow{D^{\mu}} H = H^{\dagger} \sigma^a D^{\mu} H - \bigl( D^{\mu} H \bigr)^{\dagger} \sigma^a H$.  
One can also consider the operator $\bigl( H^{\dagger} \sigma^a \overset\leftrightarrow{D^{\mu}} H \bigr)^2$, but this one can be related to $\Ocal_{2W}$ below using the field equations \pref{eq:field_eqns}, and therefore we do not consider it separately.  
There are three operators involving the isospin gauge field alone, 
\begin{subequations}\label{eq:op_W}
\begin{align}
	\Ocal_{2W} & = - \frac{1}{2} \bigl( (\Dhat^{\mu} W_{\mu \nu})^a \bigr)^2 \\
	\Ocal_{3W} & = \frac{1}{3!} g \, \epsilon^{abc} W_{\mu}^{a \, \nu} W_{\nu \rho}^{b} W^{c \, \rho \mu} \\
	\Ocal_{3\tilde{W}} & = \frac{1}{3!} g \, \epsilon^{abc} \widetilde{W}_{\mu}^{a \, \nu} W_{\nu \rho}^{b} W^{c \, \rho \mu} 
\end{align}
\end{subequations}
where 
$\Dhat^{ac}_{\mu} = \delta^{ac} \partial_{\mu} + g \, \epsilon^{abc} W^{b}_{\mu}$ is the covariant derivative for a field that transforms in the adjoint representation of $\SU{2}$ [and singlet representations of $\U{1}_Y$ and $\SU{3}_c$], and specifically $(\Dhat^\mu W_{\mu \nu})^a = \Dhat^{ac}_{\mu} W^{c \, \mu \nu}$.  
There is $1$ operator involving only the hypercharge gauge field: 
\begin{subequations}\label{eq:op_B}
\begin{align}
	\Ocal_{2B} & = - \frac{1}{2} \bigl( \partial^{\mu} B_{\mu \nu} \bigr)^2 
	\per
\end{align}
\end{subequations}
The remaining operators can be broken into a group of five that vanish if the isospin field is in vacuum ($W_{\mu \nu}^{a} = 0$) 
\begin{subequations}\label{eq:op_HW}
\begin{align}
	\Ocal_W & = \frac{ig}{2} \bigl( H^{\dagger} \sigma^a \overset\leftrightarrow{D^{\mu}} H \bigr) ( \Dhat^{\nu} W_{\mu \nu} )^a \\
	\Ocal_{WW} & = g^2 \, H^{\dagger} H \, W_{\mu \nu}^{a} W^{a \, \mu \nu} \\
	\Ocal_{W\tilde{W}} & = g^2 \, H^{\dagger} H \, W_{\mu \nu}^{a} \widetilde{W}^{a \, \mu \nu} \\
	\Ocal_{HW} & = i g \, \bigl( D^{\mu} H \bigr)^{\dagger} \sigma^{a} \bigl( D^{\nu} H \bigr) W_{\mu \nu}^{a} \\
	\Ocal_{H\tilde{W}} & = i g \, \bigl( D^{\mu} H \bigr)^{\dagger} \sigma^{a} \bigl( D^{\nu} H \bigr) \widetilde{W}_{\mu \nu}^{a} 
	\com
\end{align}
\end{subequations}
a second group of five that vanishes if the hypercharge field is in vacuum ($B_{\mu \nu} = 0$) 
\begin{subequations}\label{eq:op_HB}
\begin{align}
	\Ocal_B & = \frac{ig^{\prime}}{2} \bigl( H^{\dagger} \overset\leftrightarrow{D^{\mu}} H \bigr) \partial^{\nu} B_{\mu \nu} \\
	\Ocal_{BB} & = g^{\prime 2} \, H^{\dagger} H \, B_{\mu \nu} B^{\mu \nu} \\
	\Ocal_{B\tilde{B}} & = g^{\prime 2} \, H^{\dagger} H \, B_{\mu \nu} \widetilde{B}^{\mu \nu} \\
	\Ocal_{HB} & = i g^{\prime} \, \bigl( D^{\mu} H \bigr)^{\dagger} \bigl( D^{\nu} H \bigr) B_{\mu \nu} \\
	\Ocal_{H\tilde{B}} & = i g^{\prime} \, \bigl( D^{\mu} H \bigr)^{\dagger} \bigl( D^{\nu} H \bigr) \widetilde{B}_{\mu \nu} 
	\com
\end{align}
\end{subequations}
and a group of two that vanishes if either gauge field is in vacuum 
\begin{subequations}\label{eq:op_WB}
\begin{align}
	\Ocal_{WB} & = g^{\prime} g \, H^{\dagger} \sigma^a H \, W_{\mu \nu}^{a} B^{\mu \nu} \\
	\Ocal_{W\tilde{B}} & = g^{\prime} g \, H^{\dagger} \sigma^a H \, W_{\mu \nu}^{a} \widetilde{B}^{\mu \nu} 
	\per
\end{align}
\end{subequations}
The operators involving a field with a tilde are {\it CP} odd, and we have defined $\widetilde{F}_{\mu \nu} = \frac{1}{2} \epsilon^{\mu \nu \rho \sigma} F_{\rho \sigma}$ for $F = W^a,B$ and with $\epsilon^{0123} = +1$.  
In total, we have enumerated $20$ operators.  
A set of four identities relates some of the operators to one another \cite{Elias-Miro:2013mua}, but we will not make use of these relations.  

\subsection{Constraints on Dimension-Six Operators}\label{sub:Constraints}

The dimension-six operators lead to deviations away from the SM predictions for various Higgs and electroweak processes.  
These processes have been measured at the LEP and LHC colliders to be consistent with the SM, and consequently the operator coefficients $c_i / \Lambda^2$ are constrained around zero.  
Here we briefly discuss constraints on a few of the operators, and the reader can find more details in Refs.~\cite{Elias-Miro:2013mua,Pomarol:2013zra,Falkowski:2013dza}.  
We do not impose these experimental bounds when calculating the sphaleron energy, but rather we present them here to give the reader a feel for the numbers.  

The operator $\Ocal_T$ violates the custodial isospin symmetry \cite{Logan:2014jla}, and therefore it can lead to a large deviation in the $\rho$ parameter.  
The $\rho$ parameter was measured with $0.01\%$ precision by the LEP collider to be consistent with the SM prediction, and this translates into a strong constraint on the operator coefficient:  $-(0.005/v^2) < (c_T/\Lambda^2) < (0.001/v^2)$ or equivalently $\Lambda / \sqrt{|c_T|} \gtrsim (3.5-7.8) \TeV$ at $95\%$ C.L.  

The operator $\Ocal_6$ is a modification to the Higgs potential, which only affects the self-coupling of the Higgs boson.  
Since the Higgs self-coupling is still very poorly measured, this operator coefficient is very weakly constrained.  
To get a sense of the numbers, we estimate at what value of $(c_6/\Lambda^2)$ there is an $O(1)$ deviation in the Higgs trilinear self-coupling $\lambda_{hhh}$ away from its SM prediction, which is still consistent with LHC data.  
The self-coupling is given by $\lambda_{hhh} = d^3 V / d\phi^3 |_{\phi=v} = 3 m_h^2 / v + 6 (c_6/\Lambda^2) v^3$.  
The $\Ocal_6$ correction is comparable to the SM prediction when $\Lambda / \sqrt{|c_6|} \approx \sqrt{2} v^2 / m_h \simeq 700 \GeV$.  

Various operators affect the coupling of the Higgs boson to the electroweak gauge bosons and photon.  
For instance, $\Ocal_{B\tilde{B}}$, $\Ocal_{HW}$, and $\Ocal_{HB}$ affect Higgs decay widths, $h \to \gamma \gamma$ and $h \to Z\gamma$.  
Since these decays are already one-loop suppressed in the SM, the constraints on the operator coefficients are quite strong.  
We have $-(0.0013/m_W^22) < (c_{B\tilde{B}}/\Lambda^2) < (0.0018/m_W^2)$ or equivalently $\Lambda/\sqrt{|c_{B\tilde{B}}|} \gtrsim (1.9-2.2) \TeV$, and also $-(0.016/m_W^2) < (c_{HW} - c_{HB})/\Lambda^2 < (0.009/m_W^2)$ or equivalently $\Lambda / \sqrt{|c_{HW} - c_{HB}|} \gtrsim (0.63-0.85) \TeV$.  
When combined with LEP bounds on electroweak precision observables and $\widehat{S}$ we also have $-(0.046/m_Z^2) < (c_{W}/\Lambda^2) < (0.050/m_Z^2)$, which is equivalent to $\Lambda / \sqrt{c_W} \gtrsim (0.43-1.3) \TeV$.  

\subsection{Higgs Scalar Potential}\label{sub:Potential}

Among the dimension-six operators, only $\Ocal_6$ contributes to the Higgs scalar potential.  
By collecting the nonderivative terms in the Lagrangian and parametrizing the Higgs field as $H = ( 0 \, , \, \phi / \sqrt{2} \bigr)^T$, we identify the scalar potential 
\begin{align}\label{eq:V_phi}
	V(\phi) = \Omega + \frac{1}{2} \mu^2 \phi^2 + \frac{1}{4} \lambda \, \phi^4 + \frac{1}{8} \frac{c_6}{\Lambda^2} \phi^6 
	\per
\end{align}
We have also included a cosmological constant energy density $\Omega$.  
The presence of the $\phi^6$ term allows the electroweak phase transition to be first order \cite{Grojean:2004xa,Delaunay:2007wb}, since taking $\mu^2 > 0$, $\lambda < 0$, and $c_6 > 0$ leads to a potential with a barrier.  
We identify the Higgs vacuum expectation value $v$ and Higgs boson mass $m_h^2$ by the conditions $V^{\prime}(v) = 0$ and $V^{\prime \prime}(v) = m_h^2$.  
We also fix the cosmological constant by imposing $V(v) = 0$.  
These three constraints determine the three parameters
\begin{align}\label{eq:params}
	\Omega = \frac{1}{8} m_h^2 v^2 - \frac{1}{8} \frac{c_6}{\Lambda^2} v^6 
	\ , \qquad 
	\mu^2 = - \frac{1}{2} m_h^2 + \frac{3}{4} \frac{c_6}{\Lambda^2} v^4 
	\ , \quad \text{and} \qquad 
	\lambda = \frac{m_h^2}{2v^2} - \frac{3}{2} \frac{c_6}{\Lambda^2} v^2 
	\per
\end{align}
For $c_6 = 0$ we regain the Standard Model Higgs potential.  
For $c_6 / \Lambda^2 > (2/3) \, m_h^2 / v^4$ the Higgs mass parameter becomes positive ($\mu^2 > 0$), and a barrier separates the local minimum at the origin ($\phi = 0$) from a local minimum at the electroweak vacuum ($\phi = v$).  
For $v \simeq 246.22 \GeV$ and $m_h \simeq 125.09 \GeV$ the inequality above corresponds to $\Lambda / \sqrt{c_6} < 593 \GeV$.  

\subsection{Field Equations and Stress Energy}\label{sub:Field_Eqn}

The dimension-six operators modify the equations of motion for the Higgs and isospin gauge fields.  
We derive the equations of motion from the Lagrangian in \eref{eq:L_def} by applying the principle of least action ($\delta S = 0$).  
We find the field equations to be 
\begin{subequations}\label{eq:field_eqns}
\begin{align}
	0 & = D_{\mu} D^{\mu} H + \mu^2 H + 2 \lambda \bigl( H^{\dagger} H \bigr) H + \sum_i \frac{c_i}{\Lambda^2} \, K_i^H \\
	0 & = \bigl( D^{\mu} W_{\mu \nu} \bigr)^a + \bigl( i g \frac{1}{2} \bigr) H^{\dagger} \sigma^a \overset\leftrightarrow{D_{\nu}} H  + \sum_i \frac{c_i}{\Lambda^2} \, \bigl( K_i^W \bigr)_\nu^a \\
	0 & = \partial^{\mu} B_{\mu \nu} + \bigl( i g^{\prime} \frac{1}{2} \bigr) H^{\dagger} \overset\leftrightarrow{D_{\nu}} H  + \sum_i \frac{c_i}{\Lambda^2} \, \bigl( K_i^B \bigr)_\nu 
\end{align}
\end{subequations}
where the contributions from the dimension-six operators are captured by 
\begin{subequations}
\begin{align}
	& \qquad K^H \equiv \partial_\mu \frac{\delta \Ocal}{\delta \partial_\mu H^{\dagger}} - \frac{\delta \Ocal}{\delta H^{\dagger}} \\ 
	& \qquad K^W \equiv - \partial_\alpha \partial_\beta \frac{\delta \Ocal}{\delta \partial_\alpha \partial_\beta W^{a \nu}} + \partial_\alpha \frac{\delta \Ocal}{\delta \partial_\alpha W^{a \nu}} - \frac{\delta \Ocal}{\delta W^{a \nu}} \\
	& \qquad K^B \equiv - \partial_\alpha \partial_\beta \frac{\delta \Ocal}{\delta \partial_\alpha \partial_\beta B^{\nu}} + \partial_\alpha \frac{\delta \Ocal}{\delta \partial_\alpha B^{\nu}} - \frac{\delta \Ocal}{\delta B^{\nu}} 
	\per
\end{align}
\end{subequations}
We calculate these terms for each of the $20$ operators and present the results in \tref{tab:covariant}.   
To our knowledge this is the first time that that a complete list of dimension-six modifications to the electroweak field equations has been calculated.  

\begin{landscape}
\thispagestyle{empty}

\makeatletter
\setlength{\@fptop}{-50pt}
\makeatother

\begin{table}[p]
\hspace*{-3.5cm}
\newcommand{\tabincell}[2]{\begin{tabular}{@{}#1@{}}#2\end{tabular}}
\begin{tabular}{>{$}c<{$}|>{$}c<{$}||>{$}p{4.2cm}<{$}|>{$}p{8.5cm}<{$}|>{$}p{6.5cm}<{$}}
\multicolumn{2}{c|}{$\Ocal_i$} & \multicolumn{1}{|c|}{$K_i^H$} & \multicolumn{1}{|c|}{$K_i^W$} & \multicolumn{1}{|c|}{$K_i^B$}  \\ \hline \hline
	\Ocal_H 
	& \frac{1}{2} \bigl( \partial_\mu ( H^{\dagger} H ) \bigr)^2 
	& H \partial_\mu \partial^\mu \bigl( H^{\dagger} H \bigr) 
	& 0 
	& 0 
	\\\hline
	
	\Ocal_T 
	& \frac{1}{2} \bigl( H^{\dagger} \overset\leftrightarrow{D^{\mu}} H \bigr)^2 
	&  \tabincell{c}{$-2 (D_{\mu} H) (H^{\dagger} \overset\leftrightarrow{D^{\mu}} H)$\\$- H \, \partial_{\mu} ( H^{\dagger} \overset\leftrightarrow{D^{\mu}} H )  $}
 	& ig (H^{\dagger} \sigma^a H) ( H^{\dagger} \overset\leftrightarrow{D}_{\nu} H ) 
	& ig^{\prime} (H^{\dagger} H) ( H^{\dagger} \overset\leftrightarrow{D}_{\nu} H ) 
	\\\hline 
	\Ocal_r 
	& H^{\dagger} H ( D_{\mu} H )^{\dagger} ( D^{\mu} H ) 
	&  \tabincell{c}{ $( H^{\dagger} H ) ( D_{\mu} D^{\mu} H )$\\$ + \partial_\mu ( H^{\dagger} H ) ( D^{\mu} H )$\\$ - H ( D_{\mu} H )^{\dagger} ( D^{\mu} H ) $}
	& - \frac{ig}{2} (H^{\dagger} H) ( H^{\dagger} \sigma^a \overset\leftrightarrow{D}_{\nu} H )
	& - \frac{ig'}{2} (H^{\dagger} H) ( H^{\dagger} \overset\leftrightarrow{D}_{\nu} H )
	\\\hline 
	\Ocal_6 
	& - ( H^{\dagger} H )^3 
	& 3 ( H^{\dagger} H )^2 H 
	& 0 
	& 0 
	\\\hline 
	
	
	\Ocal_{2W} 
	& - \frac{1}{2} \bigl( (\Dhat^{\mu} W_{\mu \nu})^a \bigr)^2
	& 0
	& \tabincell{l}{ $( \Dhat^{\rho}  \Dhat_{\nu} \Dhat^{\mu} W_{\rho \mu} )^a- ( \Dhat^{\alpha} \Dhat_{\alpha}  \Dhat^{\mu} W_{\nu \mu} )^a$\\ $- g \epsilon^{abc} W^b_{\nu \rho} ( \Dhat_{\mu} W^{\rho \mu} )^c$ }
	& 0 
	\\\hline 
	\Ocal_{3W} 
	& \frac{1}{3!} g \, \epsilon^{abc} W_{\mu}^{a \, \nu} W_{\nu \rho}^{b} W^{c \, \rho \mu}
	& 0
	&  -g \Dhat^{ae}_{\mu} ( \epsilon^{ebc} W^{b \, \mu \alpha} W^{c}_{\alpha \nu} )
	& 0 
	\\\hline 
	\Ocal_{3\widetilde{W}} 
	& \frac{1}{3!} g \, \epsilon^{abc} \widetilde{W}_{\mu}^{a \, \nu} W_{\nu \rho}^{b} W^{c \, \rho \mu}
	& 0
	& \tabincell{l}{ $\,- \frac{g}{6} \, (\tensor{\epsilon}{_\nu^\alpha^\varepsilon_\rho}) \Dhat^{ab}_{\alpha} (\epsilon^{bcd} W^c_{\varepsilon \mu} W^{d \rho \mu} ) $\\ $- \frac{g}{3} \Dhat^{ab}_{\alpha} (\epsilon^{bcd} W^{c}_{\nu \mu} \widetilde{W}^{d \alpha \mu} ) $\\$- \frac{g}{3} \Dhat^{ab}_{\alpha} ( \epsilon^{bcd} \widetilde{W}^{c}_{\nu \mu} W^{d \alpha \mu} )$}
	& 0 
	\\\hline 

	\Ocal_{2B} 
	& - \frac{1}{2} \bigl( \partial^{\mu} B_{\mu \nu} \bigr)^2
	& 0
	& 0
	& \partial_{\nu} \partial^{\mu} \partial^{\rho} B_{\mu \rho} - \partial^{\mu} \partial_{\mu} \partial^{\rho} B_{\nu \rho}
	\\ 
	\hline
	\Ocal_{W} 
	& \frac{ig}{2} \bigl( H^{\dagger} \sigma^a \overset\leftrightarrow{D^{\mu}} H \bigr) ( \Dhat^{\nu} W_{\mu \nu} )^a
	& \tabincell{l}{ $- \frac{ig}{2} \big[ (\sigma^a D^{\mu} H) ( \Dhat^{\nu} W_{\mu \nu} )^a$\\$ \,+ (D^{\mu} \sigma^a H ) ( \Dhat^{\nu} W_{\mu \nu} )^a$ \\$ \,+ ( \sigma^a H ) \partial^\mu ( \Dhat^{\nu} W_{\mu \nu} )^a \big] $ }
	& \tabincell{l}{
	$- \frac{g^2}{2} (H^{\dagger} H) ( \Dhat^{\mu} W_{\nu \mu} )^a- \frac{i g^2}{2} \epsilon^{abc} (H^{\dagger} \sigma^b \overset\leftrightarrow{D^{\mu}} H) W^c_{\nu \mu}$\\
	$- \frac{ig}{2} \Dhat^{ab}_{\mu} \Dhat^{bc}_{\nu} (H^{\dagger} \sigma^c \overset\leftrightarrow{D^{\mu}} H) + \frac{ig}{2} \Dhat^{ac}_{\rho}  \Dhat^{cb \rho} (H^{\dagger} \sigma^b \overset\leftrightarrow{D}_{\nu} H) $}
	& - \frac{gg^{\prime}}{2} (H^{\dagger} \sigma^a H) (\Dhat^{\mu} W_{\nu \mu})^a
	\\\hline 
	\Ocal_{WW} 
	& g^2 \, H^{\dagger} H \, W_{\mu \nu}^{a} W^{a \, \mu \nu}
	& -g^2 \, H \, W_{\mu \nu}^{a} W^{a \, \mu \nu}
	& -4 g^2 [ \partial^\mu ( H^\dagger H ) W^a_{\nu \mu} + H^\dagger H ( \Dhat^\mu W_{\nu \mu} )^a ] 
	& 0 
	\\\hline 
	\Ocal_{W\widetilde{W}} 
	& g^2 \, H^{\dagger} H \, W_{\mu \nu}^{a} \widetilde{W}^{a \, \mu \nu}
	& -g^2 \, H \, W_{\mu \nu}^{a} \widetilde{W}^{a \, \mu \nu}
	& -4 g^2 [ \partial^\mu ( H^\dagger H ) \wt{W}^a_{\nu \mu} + H^\dagger H ( \Dhat^\mu \wt{W}_{\nu \mu} )^a ] 
	& 0 
	\\\hline 
	\Ocal_{HW} 
	& ig \, ( D^{\mu} H )^{\dagger} \sigma^{a} ( D^{\nu} H ) W_{\mu \nu}^{a}
	& \tabincell{l}{ $ig ( D^{\mu} \sigma^a D^{\nu} H ) W^a_{\mu \nu} $\\$+ ig ( \sigma^a D^{\nu} H ) \partial^\mu W^a_{\mu \nu} $}
	& \tabincell{l}{ $\frac{g^2}{2} [  H^{\dagger} \sigma^a \sigma^b ( D^{\mu} H )+ ( D^{\mu} H )^{\dagger} \sigma^b \sigma^a H ] \, W^{b}_{\nu \mu}$\\$- ig \Dhat^{ab}_{\mu} \big[ ( D_{\nu} H )^{\dagger} \sigma^b D^{\mu} H- ( D^{\mu} H )^{\dagger} \sigma^b D_{\nu} H \big] $}
	& \frac{g g^{\prime}}{2} [ H^{\dagger} \sigma^a D^{\mu} H + (D^{\mu} H)^{\dagger} \sigma^a H ] \, W^a_{\nu \mu} 
	\\\hline 
	\Ocal_{H\widetilde{W}} 
	& ig \, \bigl( D^{\mu} H \bigr)^{\dagger} \sigma^{a} \bigl( D^{\nu} H \bigr) \widetilde{W}_{\mu \nu}^{a}
	& ig ( D^{\mu} \sigma^a D^{\nu} H ) \widetilde{W}^a_{\mu \nu} 
	& \tabincell{l}{ $\frac{g^2}{2} [  H^{\dagger} \sigma^a \sigma^b ( D^{\mu} H ) + ( D^{\mu} H )^{\dagger} \sigma^b \sigma^a H ] \, \wt{W}^{b}_{\nu \mu} $\\$- i g \, \epsilon_{\nu \mu \alpha \beta} \Dhat^{ab \mu} \big[ ( D^{\alpha} H )^{\dagger} \sigma^b ( D^{\beta} H ) \big] $ }
	& \frac{g g^{\prime}}{2} [ H^{\dagger} \sigma^a D^{\mu} H + (D^{\mu} H)^{\dagger} \sigma^a H ] \, \widetilde{W}^a_{\nu \mu} 
	\\\hline 
	
	\Ocal_{B} 
	& \frac{ig^{\prime}}{2} \bigl( H^{\dagger} \overset\leftrightarrow{D^{\mu}} H \bigr) \partial^{\nu} B_{\mu \nu}
	& -ig^{\prime} ( D^{\mu} H ) \partial^{\nu} B_{\mu \nu}
	& -\frac{g g^{\prime}}{2} (H^{\dagger} \sigma^a H) \partial^{\mu}  B_{\nu \mu}
	& \tabincell{l}{ $- \frac{g^{\prime 2}}{2} ( H^{\dagger} H ) \partial^{\mu} B_{\nu \mu} $\\$- \frac{ig^{\prime}}{2} [ \partial_{\mu} \partial_{\nu} ( H^{\dagger} \overset\leftrightarrow{D^{\mu}} H ) - \partial_{\mu} \partial^{\mu} ( H^{\dagger} \overset\leftrightarrow{D}_{\nu} H ) ] $}
	\\\hline 
	\Ocal_{BB} 
	& g^{\prime 2} \, H^{\dagger} H \, B_{\mu \nu} B^{\mu \nu}
	& -g^{\prime 2} \, H \, B_{\mu \nu} B^{\mu \nu}
	& 0
	& -4 g^{\prime 2} [ \partial^\mu ( H^{\dagger}H ) B_{\nu \mu} + ( H^{\dagger}H ) \partial^{\mu} B_{\nu \mu} ] 
	\\\hline 
	\Ocal_{B\widetilde{B}} 
	& g^{\prime 2} \, H^{\dagger} H \, B_{\mu \nu} \widetilde{B}^{\mu \nu}
	& -g^{\prime 2} \, H \, B_{\mu \nu} \widetilde{B}^{\mu \nu}
	& 0
	&  -4 g^{\prime 2} \partial^{\mu} ( H^{\dagger} H ) \widetilde{B}_{\nu \mu} 
	\\\hline 
	\Ocal_{HB} 
	& i g^{\prime} ( D^{\mu} H )^{\dagger} ( D^{\nu} H ) B_{\mu \nu}
	& \tabincell{l}{ $ i g^{\prime} ( D^{\mu} D^{\nu} H ) B_{\mu \nu}$\\$- i g^{\prime} ( D^{\mu} H ) \partial^{\nu} B_{\mu \nu} $}
	& \frac{g g^{\prime}}{2} [ H^{\dagger} \sigma^a D^{\mu} H + ( D^{\mu} H )^{\dagger} \sigma^a H ] B_{\nu \mu} 
	& \tabincell{l}{ $\frac{g^{\prime 2}}{2} [ H^{\dagger} D^{\mu} H + (D^{\mu} H)^{\dagger} H ] B_{\nu \mu}$\\$+ ig^{\prime} \, \partial^{\mu} [ (D_{\mu} H)^{\dagger} (D_{\nu} H) - (D_{\nu} H)^{\dagger} (D_{\mu} H) ] $}
	\\\hline 
	\Ocal_{H\widetilde{B}} 
	& i g^{\prime} ( D^{\mu} H )^{\dagger} ( D^{\nu} H ) \widetilde{B}_{\mu \nu}
	& ig^{\prime} \bigl( D^{\mu} D^{\nu} H \bigr) \widetilde{B}_{\mu \nu}
	& \frac{g g^{\prime}}{2} [ H^{\dagger} \sigma^a D^{\mu} H + ( D^{\mu} H )^{\dagger} \sigma^a H ] \widetilde{B}_{\nu \mu} 
	& \tabincell{l}{ $\frac{g'^2}{2} \widetilde{B}_{\nu \mu} [ H^{\dagger} D^{\mu} H + (D^{\mu} H)^{\dagger} H ] $\\$+ ig^{\prime} \epsilon_{\mu \nu \rho \sigma}  \partial^{\mu} [ (D^{\rho} H)^{\dagger} (D^{\sigma} H) ] $}
	\\ 
	\hline
	\Ocal_{WB} 
	& g g^{\prime} \, H^{\dagger} \sigma^a H \, W_{\mu \nu}^{a} B^{\mu \nu}
	& -g g^{\prime} \, \sigma^a H \, W_{\mu \nu}^{a} B^{\mu \nu}
	& - 2 g g^{\prime} [ ( H^{\dagger} \sigma^a H ) \partial_{\mu} \tensor{B}{_\nu^\mu} + \Dhat^{ab}_{\mu} ( H^{\dagger} \sigma^b H ) \tensor{B}{_\nu^\mu} ] 
	& 2 g g^{\prime} \, \partial^\mu [ ( H^{\dagger} \sigma^a H ) \, W^a_{\mu \nu} ]
	\\\hline 
	\Ocal_{W\widetilde{B}} 
	& g g^{\prime} \, H^{\dagger} \sigma^a H \, W_{\mu \nu}^{a} \widetilde{B}^{\mu \nu}
	& -g g^{\prime} \, \sigma^a H \, W_{\mu \nu}^{a} \widetilde{B}^{\mu \nu}
	& - 2 g g^{\prime} [ ( H^{\dagger} \sigma^a H ) \partial_{\mu} \tensor{\wt{B}}{_\nu^\mu} + \Dhat^{ab}_{\mu} ( H^{\dagger} \sigma^b H ) \tensor{\wt{B}}{_\nu^\mu} ] 
	& 2 g g^{\prime} \, \partial^\mu [ ( H^{\dagger} \sigma^a H ) \, \widetilde{W}^a_{\mu \nu} ]
\end{tabular}
\caption{\label{tab:covariant}
For each of the $20$ dimension-six operators $\Ocal_i$ we calculate the modification to the field equations for the Higgs field $K_i^H$, the isospin gauge field $K_i^W$, and hypercharge gauge field $K_i^B$.  
}
\end{table}%

\end{landscape}

The dimension-six operators also modify the stress-energy tensor.  
Let us focus on the stress-energy that arises from only the Higgs and electroweak gauge fields.  
The corresponding stress-energy tensor can be written as 
\begin{align}\label{eq:T_def}
	T^{\mu \nu} \bigr|_{\rm EW} & = \left( D_{\alpha} H \right)^{\dagger} \left( D_{\beta} H \right) \bigl[ g^{\mu \alpha} g^{\nu \beta} + g^{\nu \alpha} g^{\mu \beta} - g^{\alpha \beta} g^{\mu \nu} \bigr] \nonumber \\
	& \quad - \bigl( B_{\alpha \beta} B_{\gamma \delta} + W_{\alpha \beta}^a W_{\gamma \delta}^{a} \bigr) \bigl[ g^{\alpha \mu} g^{\beta \delta} g^{\gamma \nu} - g^{\alpha \gamma} g^{\beta \delta} g^{\mu \nu} / 4 \bigr]
	+ \sum_i \frac{c_i}{\Lambda^2} \, \Tcal_i^{\mu \nu}
\end{align}
where a total divergence has been added to ensure that $T^{\mu \nu}$ is symmetrized and gauge invariant.  
The contribution from the dimension-six operators is parametrized by 
\begin{align}\label{eq:Tcal_def}
	\Tcal_i^{\mu \nu} 
	\equiv - \frac{2}{\sqrt{-g}} \frac{\delta \bigl( \sqrt{-g} \, \Ocal_i \bigr)}{\delta g_{\mu \nu}} 
	= - 2 \frac{\delta \Ocal_i}{\delta g_{\mu \nu}} - \Ocal_i \, g^{\mu \nu} 
	\per
\end{align}
We have calculated $\Tcal_i^{\mu \nu}$ for each of the $20$ dimension-six operators, and we present our results in \aref{app:EM_Tensor}.  

\subsection{Sphaleron Parametrization}\label{sub:Sphaleron}

The Weinberg-Salam model is regained by taking the $\Lambda \to \infty$ limit of the Lagrangian in \eref{eq:L_def}.  
The corresponding field equations \pref{eq:field_eqns} admit a static solution that asymptotes to a vacuum field configuration at spatial infinity, and which is a saddle point of the energy functional.  
This solution was discovered in 1983 \cite{Manton:1983nd,Klinkhamer:1984di} (see also \cite{Dashen:1974ck}) and given the name ``sphaleron,'' which is Greek for ``ready to fall.''  
The sphaleron configuration lies at a midpoint along a path in configuration space that interpolates between topologically distinct vacuum configurations.  

The sphaleron is frequently studied in the $g^\prime \to 0$ limit of the Standard Model Lagrangian.  
For $g^\prime = 0$ the hypercharge gauge field decouples, see \eref{eq:field_eqns}, and the minimum energy solution has the hypercharge gauge field in vacuum, $B_{\mu \nu} = 0$.  
Moreover, the electroweak sector of the Standard Model enjoys a custodial $\SU{2}$ symmetry under which the three weak gauge bosons transform as a degenerate, massive triplet \cite{Logan:2014jla}.  
Consequently, the sphaleron configuration has an $\SO{3}$-symmetric energy density and vanishing vector properties ({\it e.g.}, magnetic moment and angular momentum), which would select a preferred direction and thereby violate the custodial symmetry\footnote{To our knowledge, the connection between the custodial symmetry and the isotropy of the ($g^\prime = 0$) sphaleron has not been recognized previously.  Since the Standard Model Yukawa interactions also violate the custodial symmetry, radiative effects are expected to break the $\SO{3}$ symmetry of the sphaleron.  }.  
For the measured value of the weak mixing angle, $\sin^2 \theta_W = g^{\prime 2} / (g^2 + g^{\prime 2}) \simeq 0.23$, the sphaleron is anisotropic, because the hypercharge gauge field $B_{\mu}$ selects a preferred direction, which corresponds to the orientation of the sphaleron's magnetic moment \cite{Manton:1983nd,Klinkhamer:1984di,Klinkhamer:1990fi,Kleihaus:1991ks,Kunz:1992uh}.  

For simplicity, we take $g^{\prime} = 0$, which lets us consistently set $B_{\mu} = 0$.  
This vanishes some of the dimension-six operators, and we are left with only the $12$ operators appearing in Eqs.~(\ref{eq:op_H}),~(\ref{eq:op_W}),~and~(\ref{eq:op_HW}).  
As compared with the Standard Model, taking $g^{\prime} = 0$ introduces an error in the (zero-temperature) sphaleron energy calculation that is parametrically $\delta E_{{\rm sph},0} / E_{{\rm sph},0} \sim \sin^2 \theta_W$, and that evaluates to $\simeq 1 \%$ when all the numerical factors are included \cite{Klinkhamer:1984di,Klinkhamer:1990fi,Kleihaus:1991ks,Kunz:1992uh}.  
It is a natural extension of our work to retain the hypercharge gauge field and calculate the sphaleron solution at the physical value of the weak mixing angle.  

When the hypercharge gauge field is dropped ($B_{\mu} = 0$), the electroweak sphaleron can be parametrized in terms of two spherically symmetric profile functions.  
We work in the temporal gauge, $W_0^a = 0$, such that only the spatial components of the isospin gauge field are nonzero.   
In polar coordinates, we define the spatial radial coordinate $r = \sqrt{x^i x^i}$ and the 3-vector $n_i = x^i / r$.  
Then the sphaleron {\it Ansatz} can be written as (see \aref{app:Ansatz})
\begin{subequations}\label{eq:ansatz}
\begin{align}
	H(r) & = \frac{v}{\sqrt{2}} \, h(r) \, i n_a \sigma^a \begin{pmatrix} 0 \\ 1 \end{pmatrix} \\
	W_i^a(r) & = \frac{2}{g} \, \epsilon^{aij} n_j \frac{f(r)}{r} 
	\per 
\end{align}
\end{subequations}
where $h(r)$ and $f(r)$ are the dimensionless Higgs and isospin gauge field profile functions.  
We require the profile functions to satisfy the boundary conditions 
\begin{align}\label{eq:boundary}
	h(r=0) = 0 
	\ , \quad 
	f(r=0) = 0 
	\ , \quad 
	h(r\to \infty) = 1 
	\ , \quad \text{and} \quad 
	f(r\to \infty) = 1 
	\per
\end{align}
The first two conditions ensure that $H$ and $W^a_i$ are regular at the origin.  
The second two conditions ensure that $H$ and $W^a_i$ asymptote to a pure-gauge configuration at spatial infinity.  
This can be seen explicitly by writing $i n_a \sigma^a = U$ and $\sigma^a \epsilon^{aij} n_j / r = -i \partial_i U \, U^{-1}$ where $U = {\rm exp}[i \pi n_a \sigma^a / 2]$ is an element of $\SU{2}$.  
Then at spatial infinity $H \to (v/\sqrt{2}) \, U \, (0 \, , \, 1)^T$ and $ig (\sigma^a/2) W_i^a \to \partial_i U \, U^{-1}$.  

The sphaleron {\it Ansatz} \pref{eq:ansatz} lets us reduce the Higgs and isospin gauge field equations in \eref{eq:field_eqns} to simple one-dimensional equations of motion for the two profile functions.  
We find 
\begin{subequations}\label{eq:eqn_of_motion}
\begin{align}
	\frac{d}{d\xi} \Bigl( \xi^2 \frac{dh}{d \xi} \Bigr) & = 2 (1-f)^2 h + \frac{\mu^2}{g^2 v^2} \xi^2 h + \frac{\lambda}{g^2} \xi^2 h^3 + g^2 v^2 \sum_i \frac{c_i}{\Lambda^2} \, k_i^H  \\
	\xi^2 \frac{d^2f}{d \xi^2} & = 2 f (1-f) (1-2f) - \frac{\xi^2}{4} (1-f) h^2 + g^2 v^2 \sum_i \frac{c_i}{\Lambda^2} \, k_i^W 
\end{align}
\end{subequations}
where $\xi = gvr$ is a dimensionless radial coordinate.  
In the Standard Model one can write $\mu^2 = - \lambda v^2$ and collect the terms arising from the Higgs scalar potential into $(\lambda/g^2) \xi^2 (h^2-1) h$.  
However, the $H^6$ operator leads to a different relation among the parameters, which was given by \eref{eq:params}.  
The contribution from dimension-six operators is parametrized by $k_i^H$ and $k_i^W$, which can be derived as 
\begin{subequations}
\begin{align}
	k_i^H & = \left( \frac{d}{d\xi} \frac{\delta}{\delta \, dh/d\xi} - \frac{\delta}{\delta h} \right) \left( \frac{\xi^2}{g^4 v^6} \Ocal_i \right) \\ 
	k_i^W & = \frac{\xi^2}{8} \left( - \frac{d^2}{d\xi^2} \frac{\delta}{\delta \, d^2f/d\xi^2} + \frac{d}{d\xi} \frac{\delta}{\delta \, df/d\xi} - \frac{\delta}{\delta f} \right) \left( \frac{\xi^2}{g^4 v^6} \Ocal_i \right) 
	\per
\end{align}
\end{subequations}
We have calculated these terms, and our results are summarized in \tref{tab:sphaleron}.  

Let us pause to discuss the operator $\Ocal_T$.  
Among all the operators in \tref{tab:sphaleron}, only $\Ocal_T$ fails to respect the $\SO{3}$ symmetry, but rather it respects only an $\SO{2}$ symmetry.  
This is because $\Ocal_T$ violates the custodial $\SU{2}$ isospin symmetry \cite{Logan:2014jla}.  
Our $\SO{3}$-symmetric {\it Ansatz} \pref{eq:ansatz} will not give the correct sphaleron solution when $\Ocal_T$ is included.  
Instead we should work with an axially symmetric {\it Ansatz}, such as the one used to study the sphaleron solution at nonzero weak mixing angle \cite{Manton:1983nd,Klinkhamer:1984di,Klinkhamer:1990fi,Kleihaus:1991ks,Kunz:1992uh}.  
This analysis is beyond the scope of our work.  
Moreover, the operator $\Ocal_T$ is tightly constrained by electroweak precision tests, as we have already discussed in \sref{sub:Constraints}, and therefore we will not consider $\Ocal_T$ in our analysis any further.  

Let us also discuss the operator $\Ocal_{2W}$.  
Since this operator involves the second derivative of the isospin gauge field, its contribution to the equation of motion involves the fourth derivative of the profile function $f$. 
This is an impediment toward obtaining the sphaleron solution, since we must now specify four boundary conditions (rather than simply two).  
However, we can make use of the SM equation of motion to write $d^4f/d\xi^4$ in terms of lower-order derivatives of $f$ and $h$.  
Since $k_i^W$ in \eref{eq:eqn_of_motion} appears suppressed by the scale of new physics $(c_i/\Lambda^2)$, the error that we introduce by using the SM equations of motion is higher order in the EFT expansion parameter, namely $O(c_i^2 / \Lambda^4)$, and since we are only working to $O(c_i / \Lambda^2)$ the two approaches are equivalent.  
In fact, even without applying the sphaleron {\it Ansatz}, we can see directly from the field equations \pref{eq:field_eqns} that $\Ocal_{2W}$ is equivalent to $\Ocal_{W}$ at $O(c_i / \Lambda^2)$ when we also vanish the hypercharge gauge field and the fermions.  
Consequently, the sphaleron energy for $\Ocal_{2W}$ can be obtained by simply scaling the result for $\Ocal_W$.  

The energy of the sphaleron configuration is calculated by inserting the sphaleron parametrization \pref{eq:ansatz} into the stress-energy tensor \pref{eq:T_def} and performing the volume integral: $E_{{\rm sph},0} = \int \! \ud^3 x \, T^{00}$.  
The sphaleron energy may be written as 
\begin{align}\label{eq:E_sph}
	E_{{\rm sph},0} 
	= \frac{4 \pi v}{g} \int_{0}^{\infty} \! \ud \xi \, \Bigl[ 
	& 4 \biggl( \frac{df}{d\xi} \biggr)^2 
	+ \frac{8}{\xi^2} f^2 \bigl( 1 - f \bigr)^2 
	+ \frac{\xi^2}{2} \biggl( \frac{dh}{d\xi} \biggr)^2 
	+ \bigl( 1-f \bigr)^2 h^2 
	\nn & \ 
	+ \frac{1}{g^2} \xi^2 \biggl( \frac{\Omega}{v^4} + \frac{1}{2} \frac{\mu^2}{v^2} h^2 + \frac{1}{4} \lambda h^4 \biggr) 
	+ g^2 v^2 \xi^2 \sum_i \frac{c_i}{\Lambda^2} \Ecal_i
	\Bigr] 
\end{align}
where $4 \pi v / g$ has the units of energy, and the integral is dimensionless.  
In the Standard Model, the terms arising from the Higgs scalar potential can be written as $\Omega / v^4 + \mu^2 h^2 / 2 v^2 + \lambda h^4 / 4 = (\lambda / 4)( h^2 - 1)^2$, which is seen by taking the $\Lambda \to \infty$ limit of \eref{eq:params}.  
The contributions from the dimension-six operators $\Ocal_i$ are parametrized by the dimensionless energy $\Ecal_i = \Tcal^{00}_{i} / (g^2 v^6) = - \Ocal_{i} / (g^4 v^6)$.  
We have calculated these energies, and the results appear in \tref{tab:sphaleron}.  

\begin{landscape}
\thispagestyle{empty}

\begin{table}[t] 
\hspace*{-2.5cm}
\begin{tabular}{>{$}c<{$}|>{$}c<{$}||>{$}p{6.0cm}<{$}|>{$}p{6.0cm}<{$}||>{$}p{6.0cm}<{$}}
& \Ocal_i & \multicolumn{1}{|c|}{$k_i^H$} & \multicolumn{1}{|c|}{$k_i^W$} & \multicolumn{1}{|c}{$\Ecal_i$} \\ \hline \hline
	\Ocal_H 
	& \frac{1}{2} \bigl( \partial_\mu ( H^{\dagger} H ) \bigr)^2 
	& - \frac{1}{g^2} \xi h ( \xi h h^{\prime \prime} + \xi h^{\prime 2} + 2 h h^{\prime} )
	& 0 
	& \frac{1}{2g^2} h^2 h^{\prime 2}
	\\ 
	\Ocal_T 
	& \frac{1}{2} \bigl( H^{\dagger} \overset\leftrightarrow{D^{\mu}} H \bigr)^2 
	& - \frac{2}{g^2} \frac{1}{\xi^2} \frac{x^2 + y^2}{1/(g^2 v^2)} (1-f)^2 h^3
 	& \frac{1}{8g^2} \frac{x^2 + y^2}{1/(g^2 v^2)} (1-f) h^4
	& - \frac{1}{2g^2} \frac{1}{\xi^4} \frac{x^2 + y^2}{1/(g^2 v^2)} (1-f)^2 h^4
	\\ 
	\Ocal_r 
	& H^{\dagger} H \bigl( D_{\mu} H \bigr)^{\dagger} \bigl( D^{\mu} H \bigr) 
	& - \frac{1}{2g^2} h [ - 4 (1-f)^2 h^2 + 2 \xi h h^\prime + \xi^2 h^{\prime 2} + \xi^2 h h^{\prime \prime} ] 
	& - \frac{1}{8g^2} \xi^2 (1-f) h^4
	& \frac{1}{4g^2} \frac{1}{\xi^2} h^2 [ 2 (1-f)^2 h^2 + \xi^2 h^{\prime 2} ]
	\\ 
	\Ocal_6 
	& - \bigl( H^{\dagger} H \bigr)^3 
	& \frac{3}{4g^4} \xi^2 h^5 
	& 0 
	& \frac{1}{8 g^4} h^6 
	\\ 
	\hline
	\Ocal_{2W} 
	& - \frac{1}{2} \bigl( (\Dhat^{\mu} W_{\mu \nu})^a \bigr)^2 
	& 0
	& -\frac{1}{\xi^2} \bigl[ -8 f (1-f) (1-2f) (1 + 3f - 3 f^2) + 4 \xi (1 - 6f + 6f^2) ( 2 f^{\prime} - \xi f^{\prime \prime} ) + 12 \xi^2 (1-2f) f^{\prime 2} + \xi^4 f^{\prime\prime\prime\prime} \big]
	& - \frac{4}{\xi^6} [ - 2 f (1-f) (1-2f) + \xi^2 f^{\prime \prime} ]^2
	\\ 
	\Ocal_{3W} 
	& \frac{1}{3!} g \, \epsilon^{abc} W_{\mu}^{a \, \nu} W_{\nu \rho}^{b} W^{c \, \rho \mu}
	& 0
	& -\frac{2}{\xi} \bigl[ 4 f (1-f) f^{\prime} - \xi (1-2f) f^{\prime 2} - 2 \xi f (1-f) f^{\prime \prime} \bigr]
	& -\frac{16}{\xi^4} f (1-f) f^{\prime 2}
	\\ 
	\Ocal_{3\tilde{W}} 
	& \frac{1}{3!} g \, \epsilon^{abc} \widetilde{W}_{\mu}^{a \, \nu} W_{\nu \rho}^{b} W^{c \, \rho \mu}
	& 0
	& 0
	& 0
	\\ 
	\hline
	\Ocal_{W} 
	& \frac{ig}{2} \bigl( H^{\dagger} \sigma^a \overset\leftrightarrow{D^{\mu}} H \bigr) (\hat{D}^{\nu} W_{\mu \nu})^{a}
	& - \frac{4}{\xi^2} (1-f) h [ - 2 f (1-f) (1-2f) + \xi^2 f^{\prime \prime} ]
	& \frac{1}{2} [ (1-f) (1 - 7f + 8 f^2) h^2 + 2 \xi^2 h f^{\prime} h^{\prime} - \xi^2 (1-f) h^{\prime 2} + \xi^2 h^2 f^{\prime \prime} - \xi^2 (1-f) h h^{\prime \prime} ] 
	& -\frac{2}{\xi^4} (1-f) h^2 [ -2f(1-f)(1-2f) + \xi^2 f^{\prime \prime} ] 
	\\ 
	\Ocal_{WW} 
	& g^2 \, H^{\dagger} H \, W_{\mu \nu}^{a} W^{a \, \mu \nu}
	& - \frac{16}{\xi^2} h [ 2f^2 (1-f)^2 + \xi^2 f^{\prime 2} ] 
	& 2 h [ -2f (1-f) (1-2f) h + 2 \xi^2 f^{\prime} h^{\prime} + \xi^2 h f^{\prime \prime} ] 
	& - \frac{8}{\xi^4} h^2 [ 2f^2 (1-f)^2 + \xi^2 f^{\prime 2} ] 
	\\ 
	\Ocal_{W\tilde{W}} 
	& g^2 \, H^{\dagger} H \, W_{\mu \nu}^{a} \widetilde{W}^{a \, \mu \nu}
	& 0 
	& 0
	& 0
	\\ 
	\Ocal_{HW} 
	& i g \, \bigl( D^{\mu} H \bigr)^{\dagger} \sigma^{a} \bigl( D^{\nu} H \bigr) W_{\mu \nu}^{a}
	& \frac{4}{\xi^2} h [ 2 f(1-f)^3 + \xi^2 f^{\prime 2} - \xi^2 (1-f) f^{\prime\prime} ]
	& -\frac{1}{2} (1-f) [ - (1-f) (1-4f) h^2 + \xi^2 h^{\prime 2} + \xi^2 h h^{\prime \prime} ]
	& \frac{4}{\xi^4} (1-f) h [ f (1-f)^2 h + \xi^2 f^{\prime} h^{\prime} ]
	\\ 
	\Ocal_{H\tilde{W}} 
	& i g \, \bigl( D^{\mu} H \bigr)^{\dagger} \sigma^{a} \bigl( D^{\nu} H \bigr) \widetilde{W}_{\mu \nu}^{a}
	& 0
	& 0
	& 0
\end{tabular}
\caption{\label{tab:sphaleron}
Corrections to the sphaleron equations of motion \pref{eq:eqn_of_motion} that are induced by the presence of each dimension-six operator $\Ocal_i$.  In the last column, we calculate the correction to the sphaleron energy \pref{eq:E_sph}.  
}
\end{table}%
\end{landscape}

\section{Numerical Sphaleron Solution}\label{sec:Sphaleron}

The sphaleron equations of motion \pref{eq:eqn_of_motion} and the boundary conditions in \eref{eq:boundary} define a boundary value problem.  
We solve the system of equations numerically using the Newton-Kantorovich (NK) method.  
A description of the NK method can be found in \aref{app:NK_Method}.  

\begin{figure}[t]
\begin{center}
\includegraphics[height=7.5cm]{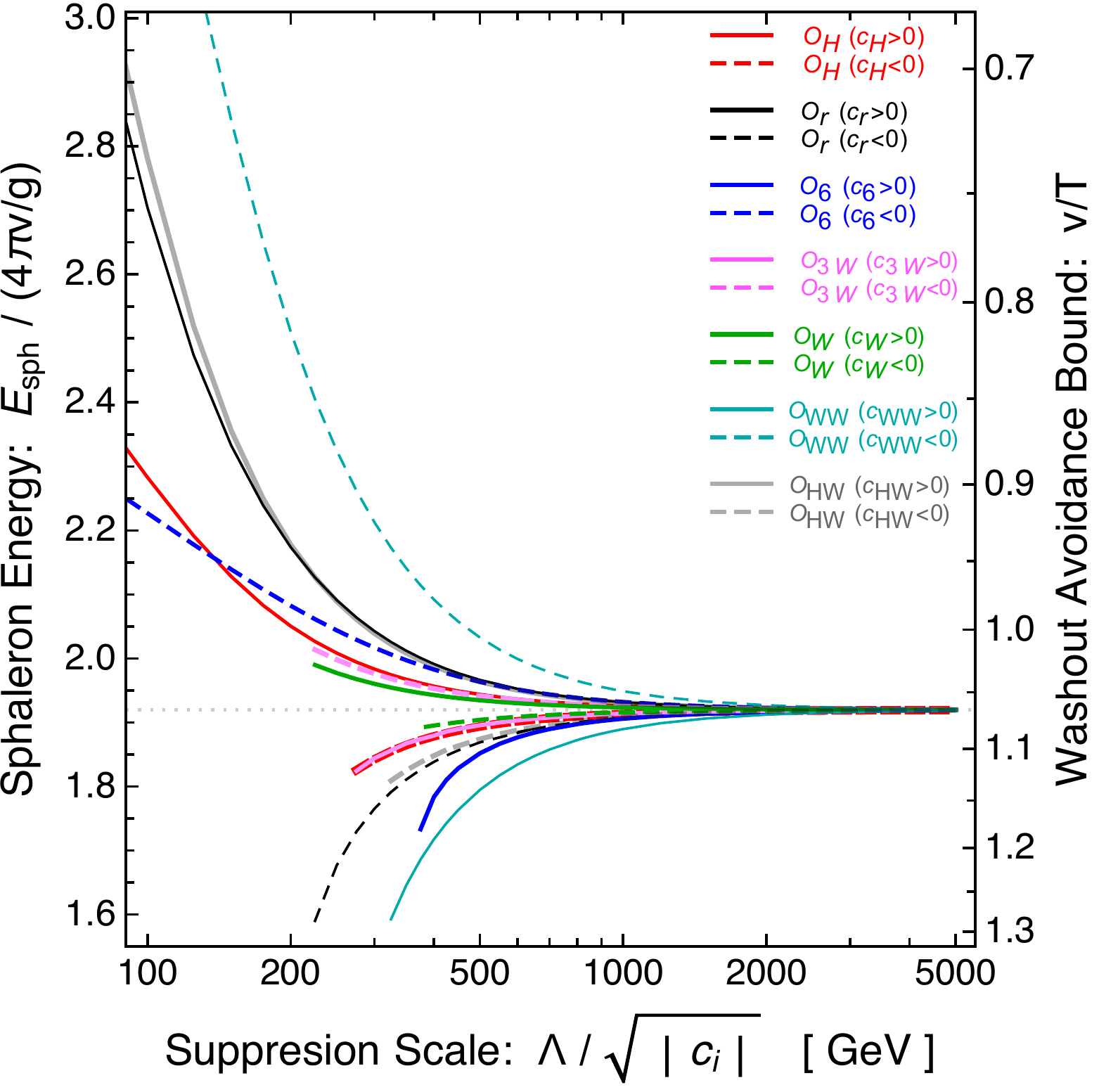} \hfill
\includegraphics[height=7.5cm]{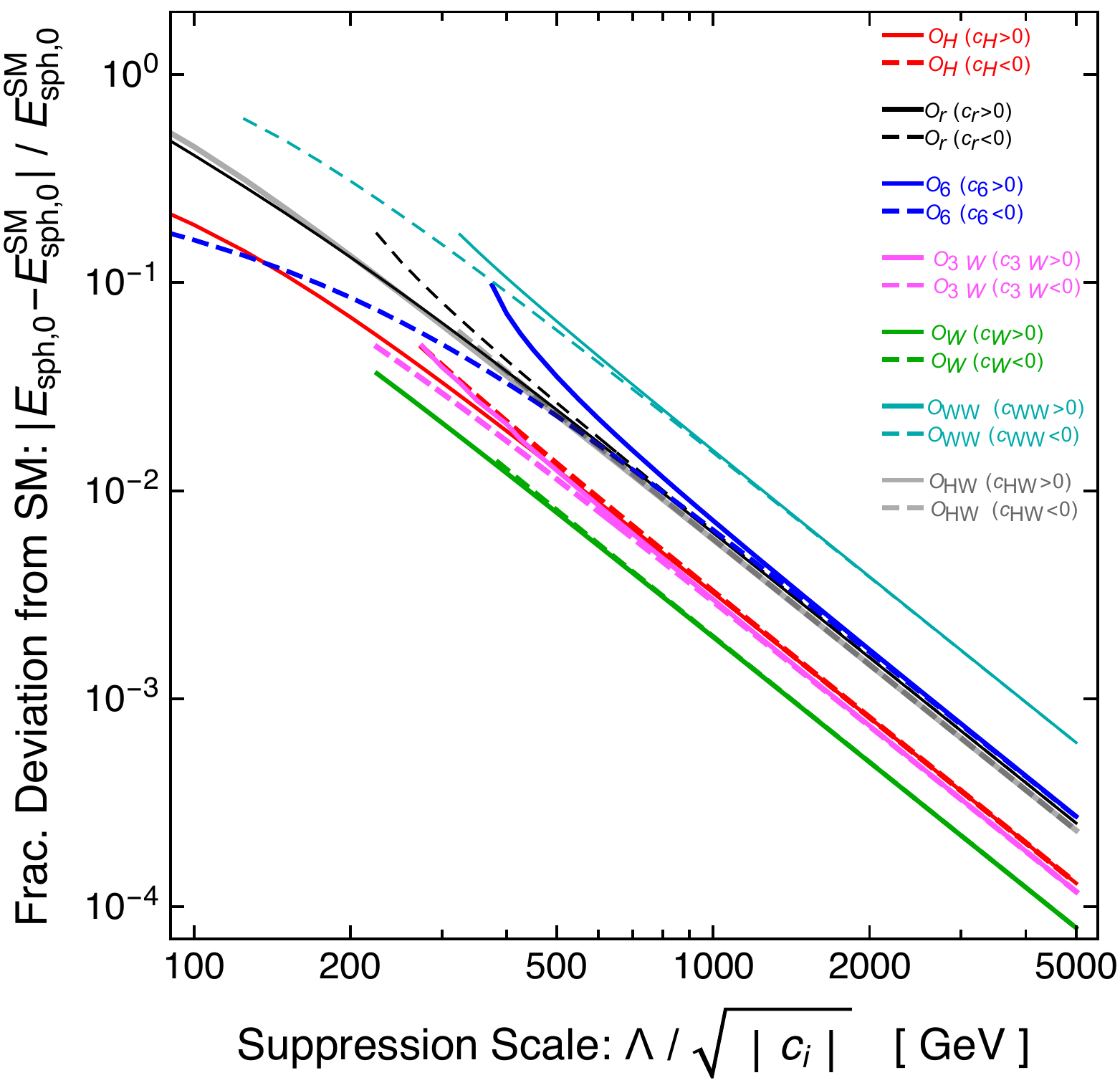} 
\caption{\label{fig:Esph}
Energy of the electroweak sphaleron in the units of $4 \pi v / g \simeq 4.7 \TeV$ as a function of the suppression scale of the dimension-six operator.  Different  colors of curves correspond to different operators as shown in the legend.  Solid (dashed) curves indicate that the operator coefficient is positive (negative).  
}
\end{center}
\end{figure}

Our main result appears in \fref{fig:Esph}.  
For a given operator $\Ocal_i$ and operator coefficient $c_i / \Lambda^2$ we calculate the sphaleron energy $E_{{\rm sph}, 0}$, which is shown on the left axis of the plot in the left panel.  
The energy is measured in units of $4\pi v/g$, and using the the measured values, $G_F = 1 / (\sqrt{2} v^2) \simeq 1.16637 \times 10^{-5} \GeV^{-2}$ and $m_W = gv/2 \simeq 80.385 \GeV$, we have $4\pi v/g = \sqrt{2} \pi / (G_F m_W) \simeq 4.738 \TeV$.  
Each curve corresponds to a different operator as indicated by the corresponding label.  
We only ``turn on'' one operator at a time, {\it i.e.} we hold all the other operator coefficients fixed to zero.  
For the solid (dashed) curves we take the operator coefficients $c_i / \Lambda^2$ to be positive (negative).  
The right axis shows the bound in \eref{eq:washout}, discussed below.  

The (zero-temperature) sphaleron energy changes as the scale of new physics is varied.  
We regain the SM result, $E_{{\rm sph},0}^\SM \simeq 1.916 \times 4\pi v/g \simeq 9.079 \TeV$, in the limit $\Lambda \to \infty$.  
As $\Lambda$ is lowered, the sphaleron energy begins to deviate from the SM prediction, and the energy can either increase or decrease depending on the operator and the sign of its coefficient.  
As we see in the right panel of \fref{fig:Esph}, the deviation away from the SM prediction scales like $1 / \Lambda^2$.  
This follows from \eref{eq:E_sph} where we can write $\Delta E_{{\rm sph},0} = (4\pi v/g) \sum_i (c_i / \Lambda^2) \int_0^\infty \ud \xi (g^2 v^2 \xi^2) \Ecal_i[h,f]$.  
Although the integral depends on $\Lambda$ implicitly through the profile functions, the dominant $\Lambda$ dependence enters through the prefactor.  
As $\Lambda \to \infty$ the profile functions approach their SM solutions, $h \to h_\SM$ and $f \to f_\SM$, and the deviation in the energy approaches $\Delta E_{{\rm sph},0} \to (4\pi v/g) \sum_i (c_i / \Lambda^2) \, g^2 v^2 \, e_i$ where $e_i \equiv \int_0^\infty \ud \xi \, \xi^2 \Ecal_i[h_\SM(\xi),f_\SM(\xi)]$.  
For each of the operators we find $e_H = 0.240$, $e_r = 0.468$, $e_{3W} = -0.223$, $e_W = 0.149$, $e_{WW} = -1.150$, and $e_{HW} = 0.436$.  For $\Ocal_6$ we define instead $e_6 \equiv \int^{\infty}_{0} \ud \xi \, \xi^2 \bigl( \Delta V \bigr) / \bigl( g^4 v^6 \, c_6/\Lambda^2 \bigr) = \int^{\infty}_{0} \ud \xi \, \xi^2  (1/8g^4) (h_\SM^2-1)^3 \simeq 0.507$ where $V(\phi)$ is the Higgs potential from \eref{eq:V_phi} and $\Delta V = V(\phi=hv) - V(\phi=hv) \bigr|_{c_6/\Lambda^2=0}$.  

As the cutoff ($\Lambda / \sqrt{|c_i|}$) is lowered, eventually the numerical (NK) method fails to obtain a solution, and this determines where we truncate the curves in \fref{fig:Esph}.  

\begin{figure}[t]
\begin{center}
\includegraphics[width=0.49\textwidth]{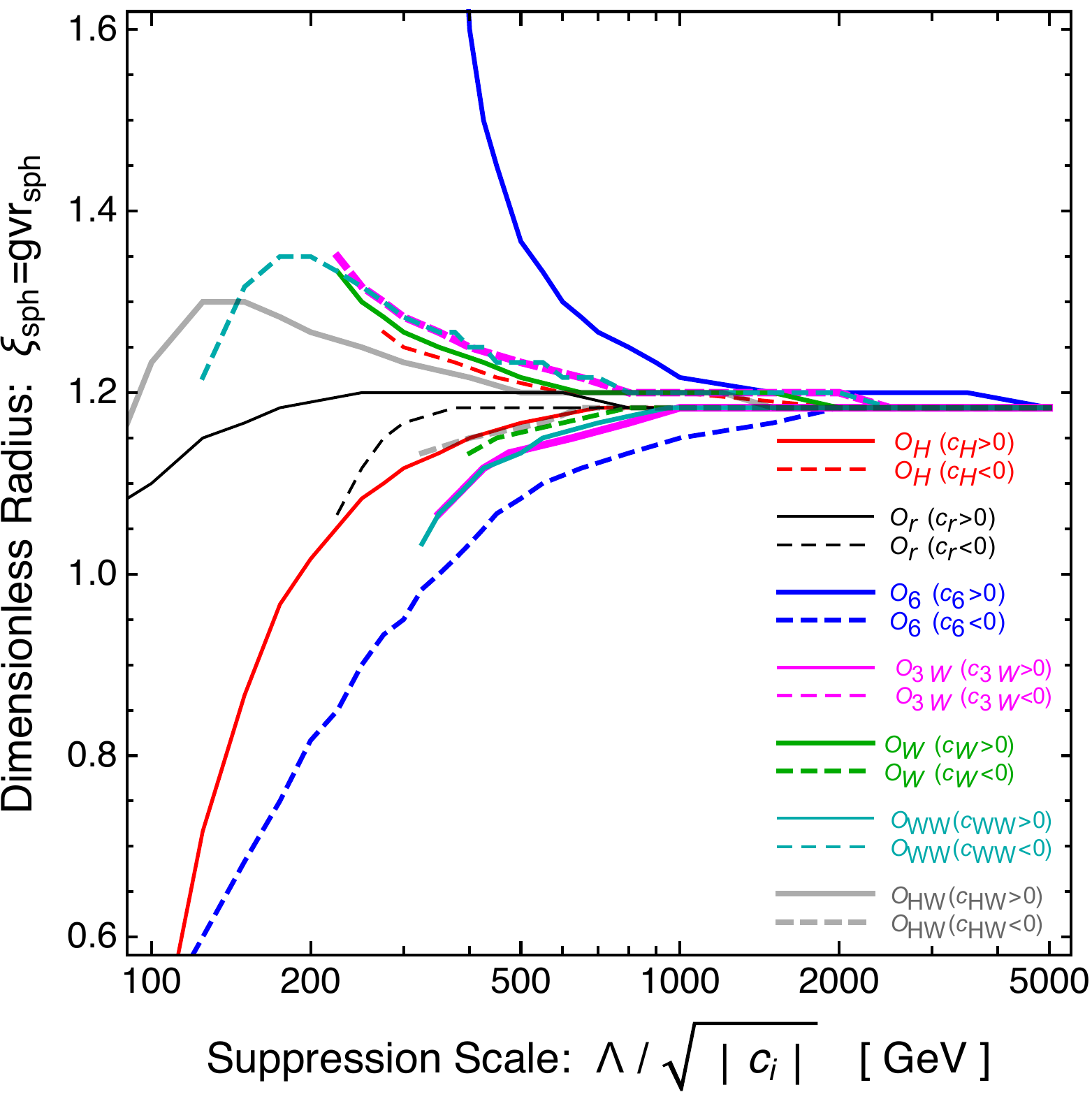} 
\caption{\label{fig:radius_E_diagram}
The dimensionless sphaleron radius as a function of the suppression scale of dimension-six operators.  }
\end{center}
\end{figure}

As we discussed in the introduction, the electroweak sphaleron leads to the washout of baryon number if it is in thermal equilibrium at the electroweak phase transition.  
To avoid this washout, the sphaleron energy must be sufficiently large compared to the temperature so that the sphaleron-mediated baryon-number-violating reactions are Boltzmann suppressed and thus out of equilibrium.  
The derivation of the washout avoidance condition appears in various references, and we review this calculation in \aref{app:Washout}.  
The result is shown to be 
\begin{align}\label{eq:washout}
	\frac{v(T)}{T} > ( 0.973 - 1.16 ) \left( \frac{E_{{\rm sph},0}}{1.916 \times 4\pi v/g} \right)^{-1} 
\end{align}
where $E_{{\rm sph},0}$ is the zero-temperature sphaleron energy and $v(T)$ is the thermal expectation value of the Higgs field at temperature $T$ in the Higgs phase (broken electroweak symmetry).  
This condition must hold both during the electroweak phase transition and at all times afterward.  
The range of values, $0.973-1.16$, reflects an $O(20\%)$ uncertainty in the calculation of the spectrum of perturbations around the sphaleron configuration (fluctuation determinant).  

On the right axis of the left panel of \fref{fig:Esph} we show the lower bound on the electroweak order parameter $v(T)/T$ from \eref{eq:washout}.  
We take the coefficient to be $1.07$, which is the arithmetic mean of $0.973$ and $1.16$.  
The washout avoidance condition can vary by as much as a few percent due to the presence of dimension-six operators while keeping $\Lambda / \sqrt{|c_i|} \gtrsim 1 \TeV$ large enough to be consistent with collider observations.  

In \fref{fig:radius_E_diagram} we show how the size of the electroweak sphaleron changes with the presence of dimension-six operators.  
To define the sphaleron size, we calculate the energy density as $\rho_{{\rm sph},0}(\xi) = (gv)^3 (4 \pi \xi^2)^{-1} d E_{{\rm sph},0} / d\xi$ with $E_{{\rm sph},0}$ given by \eref{eq:E_sph}.  Then we define a fiducial sphaleron radius $\xi_{\rm sph}$ as the width at half-maximum, $\rho_{{\rm sph},0}(\xi_{\rm sph}) = (1/2) \rho_{{\rm sph},0}(0)$.  

\section{Conclusion}\label{sec:Conclusion}

We have allowed the Standard Model to be extended with a set of $20$ dimension-six operators that are constructed from the Higgs field, isospin gauge field, and hypercharge gauge field.  
We have calculated analytically the effect of these operators on the field equations (\tref{tab:covariant}), and to our knowledge such a calculation has not appeared before in the literature.  
We have calculated analytically the sphaleron equations of motion (\tref{tab:sphaleron}) with the assumption that the hypercharge gauge field is set to zero.  
Using numerical methods, we solve the equations of motion and calculate the sphaleron energy.  

Our main result appears in \fref{fig:Esph} where we show how the sphaleron energy depends on the energy scale of new physics for each of the dimension-six operators.  
The sphaleron energy can be increased or decreased by as much as $O({\rm few} \, \%)$ as the scale of new physics is reduced to the experimental limit, roughly $\Lambda \gtrsim {\rm TeV}$.  
The lower bound on the electroweak order parameter $v(T)/T$ for the avoidance of baryon number washout \pref{eq:washout} experiences a comparable shift.  
In theories for which $v(T)/T$ depends sensitively on the model parameters, a few percent shift in the washout avoidance condition can kill or revive electroweak baryogenesis in a significant region of the parameter space.  
However, the few percent model dependence that we identify here is still subsumed by the larger $O(20\%)$ uncertainty in the washout avoidance condition that arises from imprecise calculation of the fluctuation determinant (see \aref{app:Washout}).  

There are a few places where we have made assumptions for the sake of simplicity that could be relaxed or generalized for future studies.  
As we discussed in \sref{sub:Sphaleron} we have neglected the hypercharge gauge field $B_{\mu}$ when constructing the sphaleron {\it Ansatz}.  
In the unphysical parameter regime $\sin^2 \theta_W \ll 1$, this assumption furnishes a reliable approximation to the exact sphaleron solution, but for the measured value $\sin^2 \theta_W \simeq 0.23$ the approximation begins to break down.  
While this only introduces an $O(1\%)$ error in the Standard Model, which is negligible compared to the $O(20\%)$ uncertainty in the washout condition \pref{eq:washout} for instance, the error may be larger when the dimension-six operators are added.  
It is straightforward to check this using the more general parametrizations available in Refs.~\cite{Kleihaus:1991ks,Kunz:1992uh} with the equations of motion that we have already calculated in \tref{tab:covariant}.  

In deriving the washout avoidance condition \pref{eq:washout} we assumed that the sphaleron energy at finite temperature scales linearly with the Higgs thermal expectation value \pref{Esph_scaling}.  
This is a very reliable approximation in the Standard Model, but it may break down when dimension-six operators are added since they introduce a new energy scale.  
Although we expect this effect to be a negligible $O(m_W^2/\Lambda^2)$ correction on top of what we have already calculated, it would nonetheless be interesting to test the reliability of the scaling relation by calculating the thermal sphaleron energy directly as in Refs.~\cite{Braibant:1993is,Brihaye:1993ud,Burnier:2005hp}.  
The thermal calculation introduces various subtleties associated with the presence of a plasma such as Lorentz violation by the plasma?s rest frame, thermal (Debye) mass corrections to the gauge fields, and gauge dependence that are beyond the scope of our work.

\subsubsection*{Acknowledgements}
We are grateful to Michael Fedderke for introducing us to the Newton-Kantorovich method, and we thank Kaori Fuyuto for a discussion of the washout avoidance condition calculation.  
A.J.L. is supported at the University of Chicago by the Kavli Institute for Cosmological Physics through Grant No. NSF PHY-1125897 and an endowment from the Kavli Foundation and its founder Fred Kavli.  
L.T.W. is supported by DOE grant Grant No. DE-SC0013642. 

\begin{appendix}

\section{Energy-Momentum Tensor of Dimension-Six Operators} \label{app:EM_Tensor}

Using \eref{eq:Tcal_def}, we have calculated the energy momentum tensor that arises from each of the $20$ dimension-six operators.  
In the derivation, we have used 
\begin{align}
	\frac{\delta g^{\alpha\beta}}{\delta g^{\mu\nu}} = \frac{1}{2} \Bigl( \delta_\mu^\alpha \delta_\nu^\beta + \delta_\mu^\beta \delta_\nu^\alpha \Bigr)
	\qquad \text{and} \qquad 
	\frac{\delta g^{\alpha \beta}}{\delta g_{\mu \nu}} = - \frac{1}{2} \Bigl( g^{\alpha\mu} g^{\beta\nu} + g^{\beta\mu} g^{\alpha\nu} \Bigr)
\end{align}
In this appendix we summarize our results.  

We find 
\begin{equation}\tag{22a}
T_{H}^{\mu \nu} = ( -g^{\mu \nu} g^{\alpha \beta} + 2 g^{\mu \alpha} g^{\nu \beta} ) \Big(  \frac{1}{2} \partial_{\alpha}(H^{\dagger} H)  \partial_{\beta} (H^{\dagger} H)  \Big)
\end{equation}

\begin{equation}\tag{22b}
T_{T}^{\mu \nu} = (-g^{\mu \nu} g^{\alpha \beta} + 2 g^{\mu \alpha} g^{\nu \beta})  \Big( \frac{1}{2} (H^{\dagger} \overset\leftrightarrow{D_{\alpha}} H  ) 
 (H^{\dagger} \overset\leftrightarrow{D_{\beta}}  H)  \Big)
\end{equation}

\begin{equation}\tag{22c}
T_{r}^{\mu \nu} = ( -g^{\mu \nu} g^{\alpha \beta} + g^{\mu \alpha} g^{\nu \beta} + g^{\nu \alpha} g^{\mu \beta} ) \Big( H^{\dagger} H  (D_{\alpha} H)^{\dagger} (D_{\beta} H)  \Big)
\end{equation}

\begin{equation}\tag{22d}
T_{6}^{\mu \nu} = g^{\mu \nu}  (H^{\dagger} H )^3 
\end{equation}

\begin{align}\tag{22e}
&T_{2W}^{\mu \nu} = ( -g^{\mu \nu} g^{\alpha \beta} g^{\varepsilon \rho} g^{\gamma \delta} +  2 g^{\mu \alpha} g^{\nu \beta} g^{\varepsilon \rho} g^{\gamma \delta} + 2 g^{\nu \alpha} g^{\mu \beta} g^{\varepsilon \rho} g^{\gamma \delta} + 2 g^{\alpha \beta} g^{\varepsilon \rho} g^{\mu \gamma} g^{\nu \delta}  ) \\
& \quad \quad \quad \quad \times \Big( \frac{-1}{2} \hat{D}_{\alpha}  {W^a}_{\beta \gamma}  \hat{D}_{\varepsilon} {W^a}_{\rho \delta}  \Big) \nonumber
\end{align}

\begin{align}\tag{22f}
T_{ 3W }^{\mu \nu} = ( -g^{\mu \nu} g^{\alpha \beta} g^{\varepsilon \rho} g^{\gamma \delta} + 6 g^{\mu \alpha} g^{\nu \beta} g^{\varepsilon \rho} g^{\gamma \delta} ) 
\Big( \frac{1}{3!} g \epsilon^{abc} {W^a}_{\alpha \delta} {W^b}_{\gamma \rho} {W^c}_{\varepsilon \beta}  \Big)
\end{align}

\begin{align}\tag{22g}
&T_{3 \widetilde{W}  }^{\mu \nu} = ( - g^{\mu \nu} g^{\alpha \beta} g^{\varepsilon \rho} g^{\gamma \delta} + 2 g^{\mu \alpha} g^{\nu \beta} g^{\varepsilon \rho} g^{\gamma \delta} + 2 g^{\nu \alpha} g^{\mu \beta} g^{\varepsilon \rho} g^{\gamma \delta} + 2 g^{\alpha \beta} g^{\gamma \delta} g^{\mu \rho} g^{\nu \varepsilon} ) \\
& \quad \quad \quad \quad \times \Big( \frac{1}{3!} g \epsilon^{abc} \widetilde{W}^a_{ \alpha \delta} {W^b}_{\gamma \rho} W^{c}_{\varepsilon \beta}  \Big) \nonumber
\end{align}

\begin{align}\tag{22h}
&T_{ 2B }^{\mu \nu} = ( -g^{\mu \nu} g^{\alpha \beta} g^{\gamma \delta} g^{\varepsilon \rho} + 2 g^{\mu \alpha} g^{\nu \beta} g^{\gamma \delta} g^{\varepsilon \rho}  + 
                                         2 g^{\nu \alpha} g^{\mu \beta} g^{\gamma \delta} g^{\varepsilon \rho} + 2 g^{\alpha \beta} g^{\varepsilon \rho} g^{\mu \gamma} g^{\nu \delta} )  \\
& \quad \quad \quad \quad \times \Big( \frac{-1}{2} \partial_{\alpha} B_{\beta \gamma} \partial_{\varepsilon} B_{\rho \delta} \Big) \nonumber
\end{align}

\begin{align}\tag{22i}
& T_{ W }^{\mu \nu} = ( -g^{\mu \nu} g^{\alpha \gamma} g^{\beta \delta} + g^{\mu \alpha} g^{\nu \gamma} g^{\beta \delta} + g^{\nu \alpha} g^{\mu \gamma} g^{\beta \delta} + g^{\alpha \gamma} g^{\mu \beta} g^{\nu \delta} + g^{\alpha \gamma} g^{\nu \beta} g^{\mu \delta}  ) \\
& \quad \quad \quad \quad \times \Big( \frac{ig}{2}  (H^{\dagger} \sigma^a \overset\leftrightarrow{D_{\alpha}} H) (\hat{D}_{\beta} {W^{a}}_{\gamma \delta}) \Big) \nonumber
\end{align}

\begin{align}\tag{22j}
T_{ WW }^{\mu \nu} = ( -g^{\mu \nu} g^{\alpha \gamma} g^{\beta \delta} + 4 g^{\mu \alpha} g^{\nu \gamma} g^{\beta \delta} ) \Big( g^2 (H^{\dagger} H) {W^{a}}_{\alpha \beta} {W^a}_{\gamma \delta}  \Big)
\end{align}

\begin{align}\tag{22k}
T_{ W \widetilde{W} }^{\mu \nu} = ( -g^{\mu \nu} g^{\alpha \gamma} g^{\beta \delta} + 2 g^{\mu \alpha} g^{\nu \gamma} g^{\beta \delta} 
                                                          + 2 g^{\nu \alpha} g^{\mu \gamma} g^{\beta \delta} ) \Big( g^2  (H^{\dagger} H) {W^a}_{\alpha \beta} \widetilde{W}^{a}_{\gamma \delta} \Big) \nonumber
\end{align}

\begin{align}\tag{22l}
&T_{ HW }^{\mu \nu} = ( -g^{\mu \nu} g^{\alpha \gamma} g^{\beta \delta} + g^{\mu \alpha} g^{\nu \gamma} g^{\beta \delta} + g^{\nu \alpha} g^{\mu \gamma} g^{\beta \delta} 
                                        +  g^{\alpha \gamma} g^{\mu \beta} g^{\nu \delta} + g^{\alpha \gamma} g^{\nu \beta} g^{\mu \delta} ) \\
& \quad \quad \quad \quad \times \Big( ig (D_{\alpha} H)^{\dagger} \sigma^a (D_{\beta} H)  {W^{a}}_{\gamma \delta}  \Big)  \nonumber
\end{align}

\begin{align}\tag{22m}
&T_{  H \widetilde{W} }^{\mu \nu} = ( -g^{\mu \nu} g^{\alpha \gamma} g^{\beta \delta} + g^{\mu \alpha} g^{\nu \gamma} g^{\beta \delta} + g^{\nu \alpha} g^{\mu \gamma} g^{\beta \delta} + g^{\alpha \gamma} g^{\mu \beta} g^{\nu \delta} + g^{\alpha \gamma} g^{\nu \beta} g^{\mu \delta} ) \\
& \quad \quad \quad \quad \times \Big( ig (D_{\alpha} H)^{\dagger} \sigma^a (D_{\beta} H) \widetilde{W}_{a \gamma \delta}  \Big) \nonumber
\end{align}

\begin{align}\tag{22n}
&T_{ B }^{\mu \nu} = ( -g^{\mu \nu} g^{\alpha \gamma} g^{\beta \delta} + g^{\mu \alpha} g^{\nu \gamma} g^{\beta \delta} + g^{\nu \alpha} g^{\mu \gamma} g^{\beta \delta} 
                                    + g^{\alpha \gamma} g^{\mu \beta} g^{\nu \delta} + g^{\alpha \gamma} g^{\nu \beta} g^{\mu \delta} ) \\
& \quad \quad \quad \quad \times \Big( \frac{ig'}{2} (H^{\dagger} \overset\leftrightarrow{D_{\alpha}} H) (\partial_{\beta} B_{\gamma \delta})  \Big)  \nonumber
\end{align}

\begin{align}\tag{22o}
T_{ BB }^{\mu \nu} = ( -g^{\mu \nu} g^{\alpha \gamma} g^{\beta \delta} + 4 g^{\mu \alpha} g^{\nu \gamma} g^{\beta \delta} ) \Big( g'^2 H^{\dagger} H B_{\alpha \beta} B_{\gamma \delta} \Big)
\end{align}

\begin{align}\tag{22p}
T_{ B \widetilde{B} }^{\mu \nu} = ( -g^{\mu \nu} g^{\alpha \gamma} g^{\beta \delta} + 2 g^{\mu \alpha} g^{\nu \gamma} g^{\beta \delta} + 2 g^{\nu \alpha} g^{\mu \gamma} g^{\beta \delta} ) \Big( {g'}^2 H^{\dagger} H B_{\alpha \beta} \widetilde{B}_{\gamma \delta} \Big) 
\end{align}

\begin{align}\tag{22q}
&T_{  HB }^{\mu \nu} = ( - g^{\mu \nu} g^{\alpha \gamma} g^{\beta \delta} + g^{\mu \alpha} g^{\nu \gamma} g^{\beta \delta} + g^{\nu \alpha} g^{\mu \gamma} g^{\beta \delta} + g^{\alpha \gamma} g^{\mu \beta} g^{\nu \delta} + g^{\alpha \gamma} g^{\nu \beta} g^{\mu \delta}  ) \\
& \quad \quad \quad \quad \times \Big( ig' (D_{\alpha} H)^{\dagger} (D_{\beta} H) B_{\gamma \delta} \Big) \nonumber
\end{align}

\begin{align}\tag{22r}
&T_{  H \widetilde{B} }^{\mu \nu} = ( -g^{\mu \nu} g^{\alpha \gamma} g^{\beta \delta} + g^{\mu \alpha} g^{\nu \gamma} g^{\beta \delta} + g^{\nu \alpha} g^{\mu \gamma} g^{\beta \delta} + g^{\alpha \gamma} g^{\mu \beta} g^{\nu \delta} + g^{\alpha \gamma} g^{\nu \beta} g^{\mu \delta} ) \\
& \quad \quad \quad \quad \times \Big( ig' (D_{\alpha} H)^{\dagger} (D_{\beta} H) \widetilde{B}_{\gamma \delta} \Big) \nonumber
\end{align}

\begin{align}\tag{22s}
T_{ WB  }^{\mu \nu} = ( - g^{\mu \nu} g^{\alpha \gamma} g^{\beta \delta} + 2 g^{\mu \alpha} g^{\nu \gamma} g^{\beta \delta} + 2 g^{\nu \alpha} g^{\mu \gamma} g^{\beta \delta} ) 
\Big(  g' g (H^{\dagger} \sigma^a H) {W^a}_{\alpha \beta} B_{\gamma \delta}  \Big)
\end{align}

\begin{align}\tag{22t}
T_{ W \widetilde{B}  }^{\mu \nu} = ( -g^{\mu \nu} g^{\alpha \gamma} g^{\beta \delta} + 2 g^{\mu \alpha} g^{\nu \gamma} g^{\beta \delta} + 2 g^{\nu \alpha} g^{\mu \gamma} g^{\beta \delta} ) \Big( g'g (H^{\dagger} \sigma^a H) {W^a}_{\alpha \beta} \widetilde{B}_{\gamma \delta}  \Big)
\end{align}

\setcounter{equation}{22}

\section{Sphaleron {\it Ansatz}}\label{app:Ansatz}

Here we present a more general sphaleron {\it Ansatz} and demonstrate how it reduces to \eref{eq:ansatz}.  
As in the main text, we focus here on the case $g^{\prime} = 0$ and we work in the temporal gauge $W_0^a = 0$.  
Then the Higgs and isospin gauge fields can be parametrized by five profile functions \cite{Akiba:1988ay}\footnote{We follow the notation of \rref{Spannowsky:2016ile}, which differs slightly from \rref{Akiba:1988ay}.  The correspondence between the profile functions is $F = H$, $G = K$, $A = f_A$, $B = f_B$, and $C = r^3 C$.} (see also \cite{Spannowsky:2016ile})
\begin{subequations}
\begin{align}
	H(r) & = \frac{v}{\sqrt{2}} \Bigl( \bbone \, F(r) + i n_a \sigma^a \, G(r) \Bigr) \, \begin{pmatrix} 0 \\ 1 \end{pmatrix} \\
	W_i^a(r) & = \frac{1}{g} \biggl( \epsilon^{aij} n_j \frac{1-A(r)}{r} + \bigl( \delta_{ai} - n_a n_i \bigr) \frac{B(r)}{r} + n_a n_i \frac{C(r)}{r} \biggr) 
	\per
\end{align}
\end{subequations}
An $\SU{2}_L$ gauge transformation sets $C(r) = 0$ everywhere.  
The remaining four profile functions are written as 
\begin{align}
	F = S \, \cos \phi 
	\ , \quad 
	G = S \, \sin \phi 
	\ , \quad 
	A = R \, \cos \theta 
	\ , \quad \text{and} \quad 
	B = R \, \sin \theta 
	\com 
\end{align}
and we have 
\begin{subequations}
\begin{align}
	H(r) & = \frac{v}{\sqrt{2}} \, S(r) \exp{i n_a \sigma^a \phi(r)} \, \begin{pmatrix} 0 \\ 1 \end{pmatrix} \\
	W_i^a(r) & = \frac{1}{g} \biggl( \epsilon^{aij} n_j \frac{1-R(r) \, \cos \theta(r)}{r} + \bigl( \delta_{ai} - n_a n_i \bigr) \frac{R(r) \, \sin \theta(r)}{r} \biggr) 
	\per 
\end{align}
\end{subequations}
The equations of motion for $\phi(r)$ and $\theta(r)$ impose $\phi^{\prime} = \theta^{\prime} = 0$ and $\sin \bigl( 2 \phi - \theta) = 0$, which are satisfied by taking $2\phi = \theta = \pi$.  
(We have verified that this is the case in both the SM and in the presence of the dimension-six operators.)  
Finally we write $S(r) = h(r)$ and $R(r) = 2 f(r) - 1$, and thus the {\it Ansatz} in \eref{eq:ansatz} is obtained.  

The sphaleron is often parametrized by different {\it Ans\"atze}.  
These include the Klinkhamer and Manton parametrization (1984) \cite{Klinkhamer:1984di}
\begin{subequations}
\begin{align}
	H & = \frac{v}{\sqrt{2}} \, h(\xi) \, U^{\infty} \begin{pmatrix} 0 \\ 1 \end{pmatrix} \\
\frac{\sigma^a}{2} W_i^a & = \frac{i}{g} \bigl( - f(\xi) \bigr) \partial_i U^{\infty} \, \bigl( U^{\infty} \bigr)^{-1} 
	\com
\end{align}
\end{subequations}
and the Klinkhamer and Lateveer parametrization (1990) \cite{Klinkhamer:1990fi}
\begin{subequations}
\begin{align}
	H & = \frac{v}{\sqrt{2}} \, h(\xi) \, \begin{pmatrix} 0 \\ 1 \end{pmatrix} \\
\frac{\sigma^a}{2} W_i^a & = \frac{i}{g} \bigl( 1 - f(\xi) \bigr) \bigl( U^{\infty} \bigr)^{-1} \, \partial_i U^{\infty} \
\end{align}
\end{subequations}
where in both cases 
\begin{align}
	U^{\infty} \equiv \frac{1}{r} \begin{pmatrix} z & x + i y \\ - x + i y & z \end{pmatrix} 
\end{align}
is an element of the group $\SU{2}$.  
These various parametrizations lead to identical relations for gauge-invariant quantities.

\section{Newton-Kantorovich Method}\label{app:NK_Method}

The Newton-Kantorovich method is an efficient numerical method to solve nonlinear boundary value problems. It is named after Isaac Newton and V. Kantorovich for Newton's method which can find the zeros of a real-valued function and Kantorovich's proof of the convergence of Newton's method in functional space, which is known as the Newton-Kantorovich theorem \cite{Kantorovich:1948}. From the 1950s to 1960s, Fox, Bellman, and Kalaba \cite{Fox:1957, Bellman:1965} pointed out that quasilinearization can be applied to solve two-point boundary value problems of second-order nonlinear equations numerically. The combination of the finite difference method and quasilinearization was proposed by Lee to deal with the axial diffusion in a tubular chemical reactor \cite{Lee1:1968, Lee2:1968}. The convergence, robustness, and approximate iteration steps of the NK method are discussed in Lee's article \cite{Lee2:1968}. To give a brief introduction of the NK method,  we will discuss its application of Standard Model sphaleron and calculate its energy in this section.

Here we discuss how \erefs{eq:eqn_of_motion}{eq:boundary} can be solved numerically using the NK method.  

The first step is to quasilinearize the equation of motion \pref{eq:eqn_of_motion} by substituting $h \to h + \delta h$ and $f \to f + \delta f$.  
After quasilinearization, \eref{eq:eqn_of_motion} becomes 
\begin{subequations}\label{eq:linearization}
\begin{align}
	& \xi^2 \frac{d^2 \delta h}{d \xi^2} + 2 \xi \frac{d \delta h}{d \xi} - \left[ \frac{ \mu^2}{g^2 v^2}\xi^2 + \frac{3 \lambda \xi^2}{g^2} h^2 + 2 (1-f)^2 \right] \delta h + 4 h (1-f) \delta f 
	\nn & \quad 
	= - \left[ \xi^2 \frac{d^2 h}{d \xi^2} + 2 \xi \frac{d h}{d \xi}  - \left( \frac{\mu^2}{g^2 v^2} \xi^2 h + \frac{\lambda}{g^2} \xi^2 h^3 \right) - 2 (1-f)^2 h \right] + O(\delta h^2, \delta h \, \delta f, \delta f^2) \label{eq:linearization1} \\ 
	 & \frac{\xi^2}{2} h (1-f) \delta h + \xi^2 \frac{d^2 \delta f}{d \xi^2} + \left( -12 f^2 +12f -2 - \frac{\xi^2}{4}h^2 \right) \delta f 
	 \nn & \quad 
	 = - \Big[ \xi^2 \frac{d^2 f}{d \xi^2} - 2f  (1-f) (1-2f) + \frac{\xi^2}{4} h^2 (1 -f ) \Big] +  O(\delta h^2, \delta h \, \delta f, \delta f^2) \label{eq:linearization2}
\end{align}
\end{subequations}
where the terms that are quadratic and higher order in the perturbations, $\delta h$ and $\delta f$, have been dropped.  
The right-hand sides of \erefs{eq:linearization1}{eq:linearization2} are the zeroth-order equations of motion.
After the quasilinearization the boundary conditions \pref{eq:boundary} become 
\begin{eqnarray}\label{eq:boundary:linearization}
\begin{aligned}
	&\delta h(\xi=0) = -h(\xi=0) = 0 
	\ ,  \quad
	\delta h(\xi \to \infty) = 1 - h(\xi \to \infty) = 0 
	\ ,  \\
	&\delta f(\xi=0) = -f(\xi=0) = 0 
	\ ,  \quad
	\delta f(\xi \to \infty) = 1 - f(\xi \to \infty) = 0 
	\per
\end{aligned}
\end{eqnarray}

The second step is to discretize the linearized equations of motion.  
We restrict the spatial coordinate to $\xi \in [0, L]$ and divide this range into $N$ sections uniformly.  
Then we can write $\Delta \xi = L/N$ and $\xi_i = i \Delta \xi$ for $i \in \{ 0, 1, \cdots , N \}$.  
The profile functions and their derivatives are given by 
\begin{eqnarray}\label{eq:discretization}
\begin{aligned}
 	&h(\xi = {\xi_i})  = h_i , \quad f(\xi = {\xi_i})  = f_i, \\
 	&\frac{d h}{d \xi}( \xi = {\xi_i})  = \frac{ h_{i+1} - h_{i} }{\Delta \xi}, \quad \frac{d^2 h}{d \xi^2}( \xi = {\xi_i}) = \frac{ h_{i+1}-2h_{i}+h_{i-1} }{ \Delta \xi^2 }, \\
	&\frac{d f}{d \xi}( \xi = {\xi_i})  = \frac{ f_{i+1} - f_{i} }{\Delta \xi}, \quad \frac{d^2 f}{d \xi^2}( \xi = {\xi_i}) = \frac{ f_{i+1}-2f_{i}+f_{i-1} }{ \Delta \xi^2 } ,
\end{aligned}
\end{eqnarray}
and we have similar relations for the perturbations, $\delta h$ and $\delta f$.  
The quasilinearized boundary conditions (\ref{eq:boundary:linearization}) are also discretized, which determines 
\begin{align}
	h_0 = f_0 = 1 - h_N = 1 - f_N = \delta h_0 = \delta f_0 = \delta h_N = \delta f_N = 0 \per
\end{align}
Then it is convenient to define the $(2N+2)$-dimensional array $Y = [h_0, h_1, ...,h_N, f_0, f_1, ...,f_N]$.  
With this notation, the discretized and linearized equations of motion are written concisely as 
\begin{align} \label{linearEq}
\Qcal^{(j)} \delta Y^{(j)}= q^{(j)}
\end{align}
where the $(2N+2) \times (2N+2)$-dimensional array $\Qcal^{(j)}$ and the $(2N+1)$-dimensional array $q^{(j)}$ can be determined from evaluating \eref{eq:linearization} with \eref{eq:discretization}.  
Both arrays $\Qcal^{(j)}$ and $q^{(j)}$ depend on $Y^{(j)}$.  
Since we will be solving this equation iteratively, we have added the index $j$ to denote the number of the iteration.  

As the third step, we specify a trial function to use on the zeroth iteration and solve \eref{linearEq} iteratively.  
We choose $h^{(0)}_i= \tanh(A \xi_i)$ and $f_i= \tanh(B \xi_i)$, which together form $Y^{(0)}$.  To find the numerical solution using NK method, we need to choose different sets of $A$ and $B$ for different operators and different suppression scales. For a Standard Model sphaleron, $ A = 10 $ and $ B = 3$. What should be noted is that because of the robustness of the NK method, when $A$ and $B$ are chosen properly and a physical numerical solution is found, different $A$ and $B$ will lead to the same result. 
Then $\Qcal^{(j)}$ and $q^{(j)}$ are calculated from $Y^{(j)}$, and \eref{linearEq} is solved for $\delta Y^{(j)}$.  
For the next iteration we take $Y^{(j+1)} = Y^{(j)} + \delta Y^{(j)}$.  
The convergence criterion is defined by $|\delta Y^{(j^*)}| < \varepsilon = 10^{-9}$ and the final solution is given by $Y^* = Y^{(j^*+1)}$ which can be divided into $h^*$ and $f^*$. Once the NK method has converged, we evaluate the equations of motion using the final $Y^{(j)}$ to check that the error is less than $10^{-8}$.

\begin{figure}[t]
\begin{center}
\includegraphics[height=7cm]{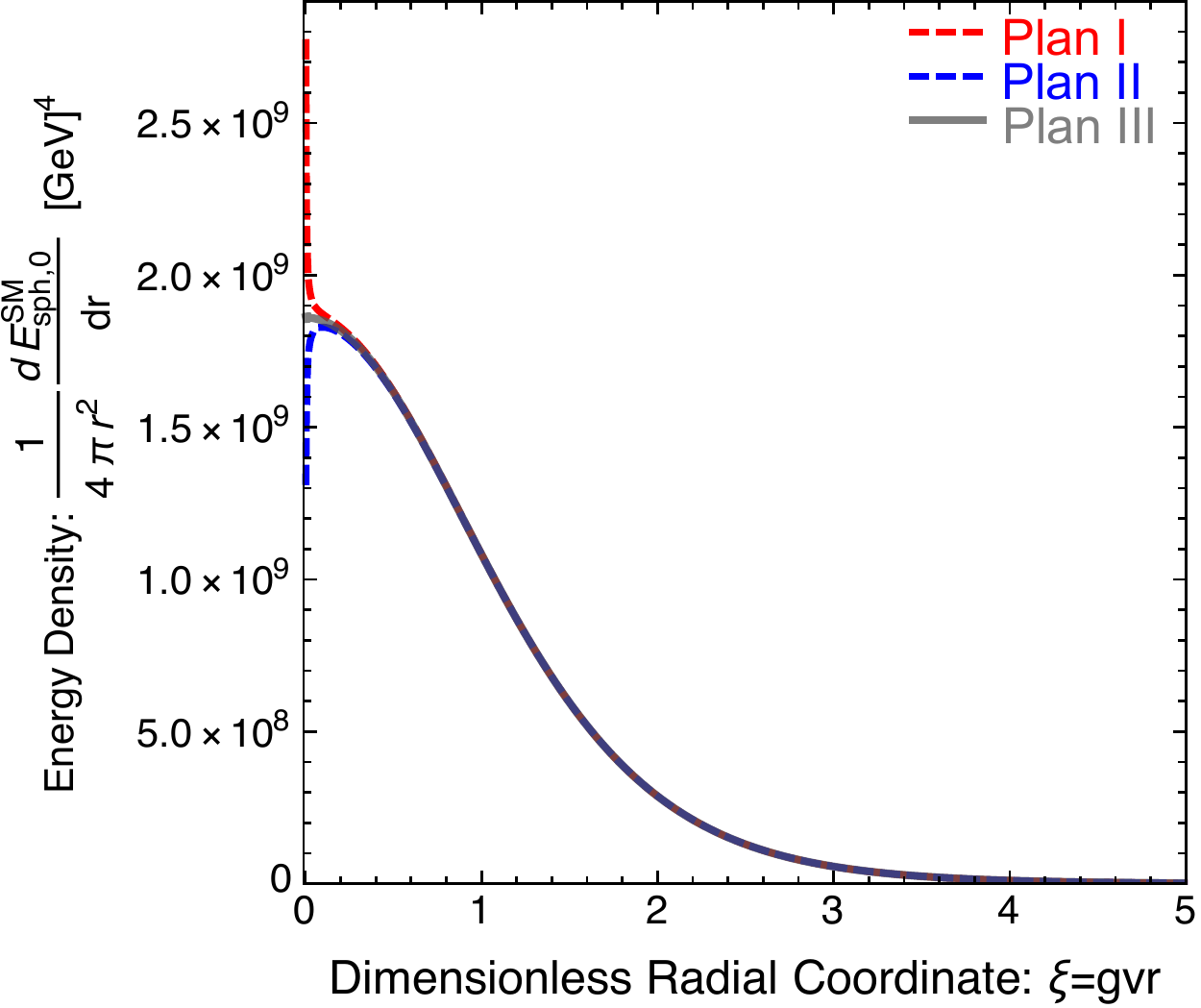} \hfill
\includegraphics[height=7cm]{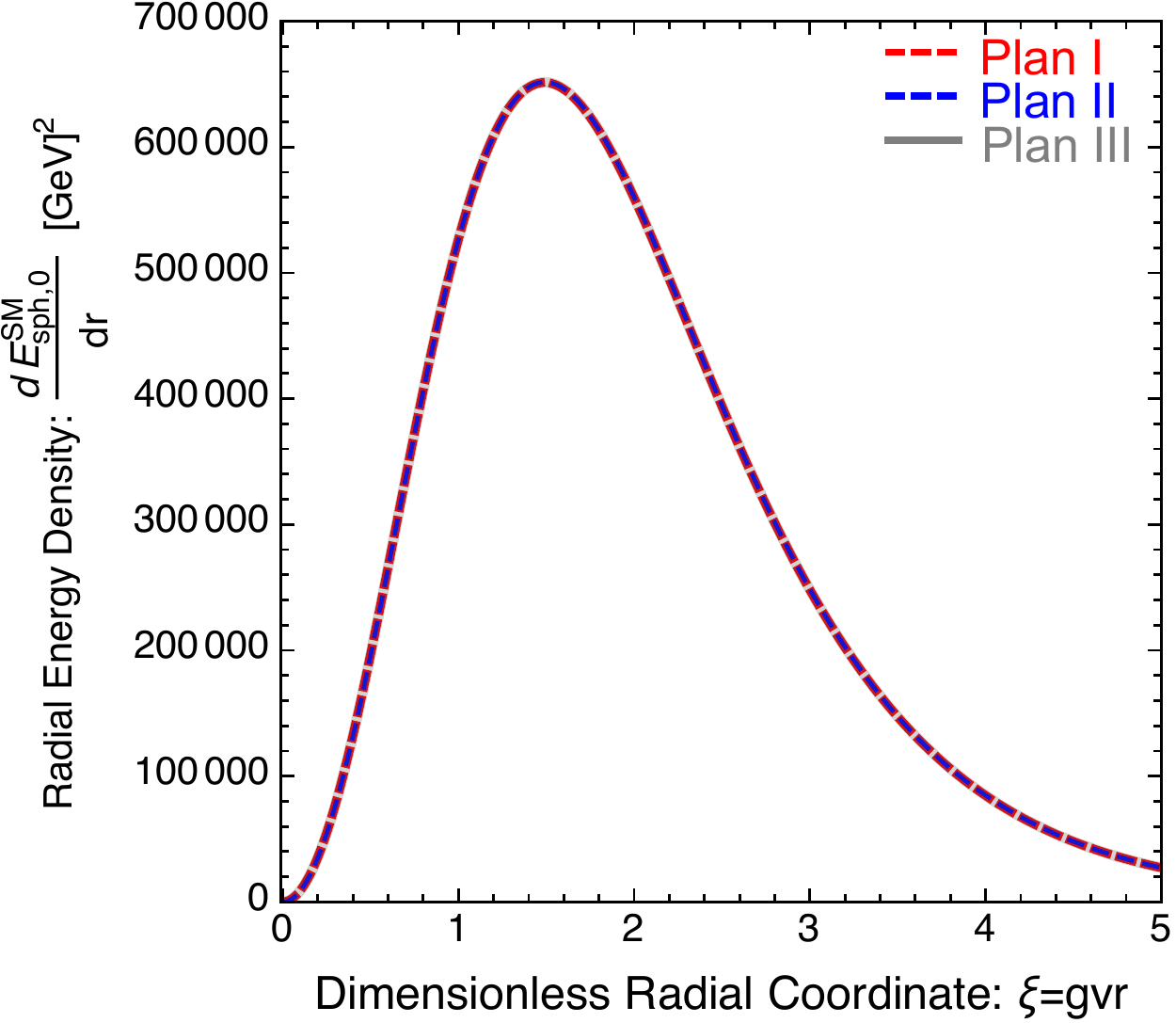} 
\caption{\label{fig:density}
The left panel is the energy density of the Standard Model sphaleron with unit ${\rm GeV}^4$, which means the Standard Model sphaleron energy per volume. The right panel is the radial energy density of the Standard Model sphaleron with unit ${\rm GeV}^2$, which means the Standard Model sphaleron energy per radial distance.
}
\end{center}
\end{figure}

Finally, the energy can be calculated using a discretized form of \eref{eq:E_sph}. However, different discretization plans of $\frac{d h^*}{d \xi}$ and $\frac{d f^*}{d \xi}$ when calculating sphaleron energy will lead to different numerical consequences, which cause algorithm uncertainty. There are mainly three plans. 
\begin{eqnarray} 
\begin{aligned}
& Plan \quad I: \quad \frac{d h^*}{d \xi} (\xi = \xi_i )= \frac{h^*_{i+1} - h^*_{i} }{\Delta \xi}, \quad  \frac{d f^*}{d \xi} (\xi = \xi_i ) =  \frac{f^*_{i+1} - f^*_{i} }{\Delta \xi} \\
& Plan \quad II: \quad \frac{d h^*}{d \xi} (\xi = \xi_i )= \frac{h^*_{i} - h^*_{i-1} }{\Delta \xi}, \quad  \frac{d f^*}{d \xi} (\xi = \xi_i ) =  \frac{f^*_{i} - f^*_{i-1} }{\Delta \xi} \\
& Plan \quad III: \quad \frac{d h^*}{d \xi} (\xi = \xi_i )= \frac{h^*_{i+1} - h^*_{i-1} }{2\Delta \xi}, \quad  \frac{d f^*}{d \xi} (\xi = \xi_i ) =  \frac{f^*_{i+1} - f^*_{i-1} }{2\Delta \xi} 
\end{aligned}
\end{eqnarray}

Plan III is more symmetric and the sphaleron energy calculated using plan III lays approximately in the middle of the energy calculated using plan I and plan II. From the left panel of \fref{fig:density} we know that using plan III we can get energy density which has a peak in the origin. We also know that using plan I we will get SM sphaleron energy a bit larger and using plan II we will get SM sphaleron energy a bit smaller so energy calculated using plan I can serve as the upper bound of SM sphaleron energy, while energy calculated using plan II can serve as the lower bound of SM sphaleron energy. In the right panel of Fig. \ref{fig:density}, curves related to these three plans nearly overlap with one another, which means when calculating total Standard Model sphaleron energy using formula $E^{SM}_{{\rm sph},0} = \int^{\infty}_{0} \ud r 4 \pi r^{2} \rho^\SM_{{\rm sph},0} $, there will be very small numerical difference.

When $N=10000$ and $L=30$, we have that $E^{I, \SM}_{{\rm sph},0} = 1.9161 (4\pi v/g)$
 and we also have that $1.9156 (4\pi v/g) \approx E^{I, \SM}_{{\rm sph},0} \leq E^{\SM}_{{\rm sph},0}\leq E^{II, \SM}_{{\rm sph},0} \approx 1.9165 (4\pi v/g)$. So we can claim that $E^{\SM}_{{\rm sph},0} = 1.9161 (4\pi v/g)$, which has algorithm uncertainty around $\Delta E \approx 0.0005  (4\pi v/g)$.  

\section{Washout Avoidance Condition}\label{app:Washout}

The sphaleron configuration is a saddle point of the energy functional with one tachyonic fluctuation mode.  
Let $\omega_-$ denote the natural frequency of the tachyonic mode.  
In general, when a system in equilibrium at temperature $T$ is prepared at the saddle point, it will decay on a time scale $\tau$ given by $\tau^{-1} = \Gamma = (\omega_- \, {\rm Im} \, F)/(\pi T)$ where $F$ is the free energy of the system evaluated at the the saddle point configuration.  
When this formula is applied to the sphaleron, the sphaleron fluctuation rate per unit volume is found to be \cite{Carson:1990jm}
\begin{align}\label{eq:Gam_sph}
	\frac{\Gamma_{\rm sph}}{V} \approx  \frac{\omega_-}{2\pi} \Ncal_{\rm tr} \Ncal_{\rm rot} \Vcal_{\rm rot} \left( \frac{\alpha_W T}{4\pi} \right)^3 \alpha_3^{-6} \kappa \exp{- E_{\rm sph}(T) / T} 
	\per
\end{align}
In the spectrum of fluctuations around the sphaleron configuration there are six zero modes, three associated with spatial translations, and three with spatial rotations.  
The factors $\Ncal_{\rm tr}$ and $\Ncal_{\rm rot}$ are normalization integrals, which can be evaluated numerically once the sphaleron profile functions are supplied.  
In the Standard Model with $\lambda/g^2 = 0.3$ they evaluate to $\Ncal_{\rm tr} \simeq 7.6$ and $\Ncal_{\rm rot} \simeq 11.2$.  
The factor $\Vcal_{\rm rot} = 8 \pi^2$ is the volume of the rotation group.  
The negative-mode frequency is evaluated numerically to be $\omega_-^2 \simeq 0.65 g^2 v(T)^2$ for $\lambda/g^2 = 0.3$ \cite{Carson:1990jm}.  
The rate picks up a factor of $1/g_3$ for each of the six zero modes where $g_3$ is the weak gauge coupling in the three-dimensional high-temperature effective theory, and we have the relations $\alpha_3 = g_3^2 / 4\pi$ and $\alpha_3 = \alpha_W T / gv(T)$.  
Using the relations above, the sphaleron decay rate can also be written as
\begin{align}
	\frac{\Gamma_{\rm sph}}{V} \approx \Bigl[ 2 \Ncal_{\rm tr} \Ncal_{\rm rot} \Vcal_{\rm rot} \frac{\omega_-}{gv(T)} \Bigr] \, T^4 \left( \frac{\alpha_W}{4 \pi} \right)^4 \left( \frac{4\pi \, v(T)}{gT} \right)^7 \kappa \exp{- E_{\rm sph}(T) / T} 
	\com
\end{align}
and the quantity in square brackets evaluates to $\simeq 1.1 \times 10^4$ for $\lambda/g^2 = 0.3$.  

The last factor in $\Gamma_{\rm sph}$ is the fluctuation determinant $\kappa$, which is defined by 
\begin{align}
	\kappa = {\rm Im} \left[ \frac{ {\rm det} \bigl( \delta^2 S_{\rm gf} / \delta \phi^2 \bigr) \bigr|_{\phi=\phi_{\rm vac}} \Delta_{\rm FP} \bigr|_{\phi = \phi_{\rm sp}} }{ {\rm det}^{\prime} \bigl( \delta^2 S_{\rm gf} / \delta \phi^2 \bigr) \bigr|_{\phi=\phi_{\rm sp}} \Delta_{\rm FP} \bigr|_{\phi = \phi_{\rm vac}} } \right]^{1/2} 
\end{align}
where $\phi$ generically represents the gauge and Higgs fields, $S_{\rm gf}$ is the gauge-fixed three-dimensional Euclidean action, $\Delta_{\rm FP}$ is the associated Faddeev-Popov determinant, and the prime indicates that zero modes are removed from the determinant.  
The calculation of $\kappa$ is the most difficult of all the terms in $\Gamma_{\rm sph}$.  
According to \rref{Carson:1990jm}, it is $\kappa \simeq e^{-2.2} \simeq 10^{-0.96} \simeq 0.11$ for $\lambda/g^2 = 0.3$.  
However the authors of \rref{Dine:1991ck} (see also \rref{Patel:2011th}) caution that the calculation of $\kappa$ requires temperature-dependent derivative terms to be included in the effective action.  
They argue that the Higgs propagator must be calculated in the background of the sphaleron configuration, and that doing so can lower the sphaleron energy by as much as $20 \%$.  
Absorbing this uncertainty into $\kappa$ implies a suppression of $e^{-0.2 E_{\rm sph}/T} \sim (1-3) \times 10^{-4}$ for $E_{\rm sph} \approx 40-45$.  
In light of these arguments, Dine {\it et. al.} \cite{Dine:1991ck} consider a range of $\kappa$ given by $10^{-4} < \kappa < 10^{-1}$.  
This range may be overly generous, and for very strongly first-order transitions ($v(T)/T > {\rm few}$) the leading-order calculation becomes more reliable.  
It should be possible to extract $\kappa$ from more recent lattice simulations of sphalerons at the Standard Model EW crossover.  
The numerical lattice results are well fit by \cite{DOnofrio:2012ni} 
\begin{align}
	\frac{\Gamma}{V} \simeq T^4 \, {\rm exp}\Bigl[ - \bigl( 147.7 \pm 1.9 \bigr) + \bigl( 0.83 \pm 0.01 \bigr) T / {\rm GeV} \Bigr] 
	\com
\end{align}
and the SM crossover occurs at $T \simeq 160 \GeV$.  
Then the uncertainty in $\Gamma$ is roughly 
\begin{align}
	\delta \frac{\Gamma}{T^4 V} \sim {\rm exp}\Bigl[ 2 \times \bigl( 1.9 + 0.01 \times 160 \bigr) \Bigr] \simeq 10^3 
	\com 
\end{align}
which is comparable to the uncertainty in $\kappa$ that we have given above.  
Suffice it to say that the calculation of the fluctuation determinant $\kappa$ is still uncertain, and we will follow \rref{Dine:1991ck} by using a wide range $10^{-4} < \kappa < 10^{-1}$ to parametrize our ignorance.  

In order for electroweak baryogenesis to be successful, the EW sphaleron transitions must be out of equilibrium in the broken phase (inside bubbles).  
Otherwise, the baryon asymmetry is washed out by a factor of 
\begin{align}
	f_{\rm w.o.} = \exp{- c_B \, X_{\rm w.o.} } 
\end{align}
where $c_B \approx (13/2) N_f = 19.5$ is an $O(1)$ number that counts the number of number of baryonic degrees of freedom \cite{Quiros:2007zz}, and 
\begin{align}
	X_{\rm w.o.} \equiv \int_{t_c}^{\infty} \ud t^{\prime} \, \Gamma_{\rm sph}
\end{align} 
where $t_c$ is the time at which the phase transition occurred.  
By changing the variable of integration, we can write 
\begin{align}
	f_{\rm w.o.} 
	= \exp{- \int_{a_c}^{\infty} \frac{\ud a^{\prime}}{a^{\prime}} \, \frac{\Gamma_{\rm sph}}{H} }
	= \exp{- \int_{0}^{T_c} \frac{\ud T^{\prime}}{T^{\prime}} \, \frac{\Gamma_{\rm sph}}{H} } 
	\per 
\end{align} 
At the time of the phase transition, the Universe is radiation dominated and the Hubble rate is given by $3 H^2 \Mpl^2 = (\pi^2/30) g_{\ast} T^4$ where $g_{\ast} \simeq 106.75$ is the number of relativistic degrees of freedom at weak-scale temperatures and $\Mpl \simeq 2.43 \times 10^{18} \GeV$ is the reduced Planck mass.  
To ensure that the baryon-preservation condition is satisfied, we impose $(\Gamma_{\rm sph}/V) < \alpha H T^3$ with $\alpha \sim 0.1$, which models the effect of the integrals above.  
This condition resolves to 
\begin{align}
	\frac{E_{\rm sph}(T)}{T} & > 
	\ln \Bigl[ 2 \Ncal_{\rm tr} \Ncal_{\rm rot} \Vcal_{\rm rot} \frac{\omega_-}{gv(T)} \Bigr] 
	+ 4 \ln \left( \frac{\alpha_W}{4 \pi} \right) 
	+ 7 \ln \left( \frac{4\pi v(T)}{gT} \right)
	+ \ln \kappa 
	\nn & \qquad 
	- \ln \alpha 
	- \frac{1}{2} \ln \left( \frac{\pi^2}{90} g_{\ast} \right)
	- \ln \frac{T}{\Mpl} 
	\per
\end{align}
Using the numerical values in the text above, this evaluates to
\begin{align}\label{eq:Esph_bound}
	\frac{E_{\rm sph}(T)}{T} > (35.9-42.8) + 7 \ln \frac{v(T)}{T} - \ln \frac{T}{100 \GeV} 
\end{align}
where the range corresponds to $\kappa = (10^{-4} - 10^{-1})$.  

Now we make an important assumption.  
In the Standard Model, the thermal energy $E_{\rm sph}(T)$ obeys an approximate scaling relation \cite{Braibant:1993is,Brihaye:1993ud}
\begin{align}\label{Esph_scaling}
	E_{\rm sph}(T) \approx E_{{\rm sph},0} \frac{v(T)}{v} 
	\per
\end{align}
This relation may break down when the SM is extended by a dimension-six operator, which brings in a new energy scale.  
Nevertheless, we expect the scaling relation to be reliable up to corrections of order $O(m_W(T)^2 / \Lambda^2)$.  
For instance, the authors of \rref{Chiang:2017nmu} identified a $5\%$ deviation from the scaling relation at a benchmark parameter point in the scalar singlet extension of the Standard Model.  
Therefore we will use \eref{Esph_scaling} for the present analysis.  
A natural extension of our work is to calculate $E_{\rm sph}(T)$ directly and impose the washout avoidance condition using \eref{eq:Esph_bound}.  

With these assumptions, the washout avoidance condition is expressed as 
\begin{align}\label{eq:vT_final}
	\frac{v(T)}{T} > \biggl( ( 0.973 - 1.16 ) + 0.190 \ln \frac{v(T)}{T} - 0.0271 \ln \frac{T}{100 \GeV} \biggr) \left( \frac{E_{{\rm sph},0}}{1.91 \times 4\pi v/g} \right)^{-1} 
	\com
\end{align}
which leads to \eref{eq:washout}; see also \rref{Fuyuto:2014yia}.  
The $20\%$ uncertainty in the value of the lower bound arises from the generous range that we have taken for the fluctuation determinant, $10^{-4} < \kappa < 10^{-1}$.  
The central value of the lower bound is logarithmically sensitive to various other model-dependent factors, namely the phase transition temperature $T$, the factor $\alpha$, and the normalization integrals $\Ncal_{\rm tr}$ and $\Ncal_{\rm rot}$.  

Let us close by remarking upon a subtlety associated with gauge dependence; see \rref{Patel:2011th} for additional details.  
The washout avoidance condition, $\Gamma_{\rm sph} \ll H$, is expressed in terms of manifestly physical rates, but the calculation of $\Gamma_{\rm sph}$ may introduce spurious and unphysical dependence on the gauge-fixing condition.  
The conventional scheme for calculating $v(T)$ consists of calculating the thermal effective potential perturbatively and defining $v(T)$ to be the field value that minimizes the potential at temperature $T$.  
If $v(T)$ is evaluated by numerically minimizing the potential, which is the conventional technique, then $v(T)$ so defined does not adhere strictly to the perturbative expansion.  
As a result $v(T)$ depends upon the gauge-fixing scheme that is used for the electroweak gauge fields; {\it e.g.}, in the renormalizable $R_{\xi}$ class of gauges, $v(T)$ depends explicitly upon $\xi$.  
Then $E_{\rm sph}(T)$ inherits this spurious gauge dependence when it is estimated using \eref{Esph_scaling}, and so too does the washout avoidance condition expressed as in \eref{eq:vT_final}.  
This issue is beyond the scope of our work, since we are primarily concerned here with calculating the gauge-invariant sphaleron energy $E_{{\rm sph},0}$, but the issue of gauge dependence should be addressed upon extending our work to finite temperature.

\end{appendix}


\end{document}